\documentclass[superscriptaddress,aps,onecolumn,12pt,floatfix,altaffilletter,showpacs,showkeys,notitlepage,nofootinbib,prd]{revtex4-2}

\pdfoutput=1

\usepackage[utf8]{inputenc}
\usepackage[dvipdf,dvips,dvipdfmx]{graphicx}
\usepackage[colorlinks=true,citecolor=blue,linkcolor=blue]{hyperref}
\usepackage{amsmath}
\usepackage{amssymb}
\usepackage{soul,xcolor}
\usepackage{epsfig}
\usepackage{epstopdf}
\usepackage{url}
\usepackage{float}
\usepackage{slashed}
\usepackage{cancel}
\usepackage{textcomp}
\usepackage{calc}
\usepackage{adjustbox}
\usepackage{mathtools}
\usepackage{gauss}
\usepackage{physics}
\usepackage{gensymb}
\usepackage{bbm}
\usepackage{esint}
\usepackage{tikz-feynman}
\usepackage{bbold}
\usepackage{tensor}
\usepackage{kbordermatrix}
	\renewcommand{\kbldelim}{(}
	\renewcommand{\kbrdelim}{)}

\usepackage{indentfirst}
\usepackage[font=small,labelfont=bf]{caption}
\usepackage{wrapfig}
\usepackage{caption}
\usepackage{comment}
\usepackage{pgfplots}
	\pgfplotsset{width=5cm,compat=1.9}
\usepackage{bm}
\usepackage{fancyhdr}
\setstcolor{red}

\allowdisplaybreaks

\catcode`\@=11
\def\lsim{\mathrel{\mathpalette\@versim<}}
\def\gsim{\mathrel{\mathpalette\@versim>}}
\def\@versim#1#2{\vcenter{\offinterlineskip
\ialign{$\m@th#1\hfil##\hfil$\crcr#2\crcr\sim\crcr } }}
\catcode`\@=12

\makeatletter
\newcommand{\raisemath}[1]{\mathpalette{\raisem@th{#1}}}
\newcommand{\raisem@th}[3]{\raisebox{#1}{$#2#3$}}
\makeatother

\newcommand{\Array}[1]{\begin{pmatrix} #1 \end{pmatrix}}
\newcommand{\Lag}{\mathcal{L}}
\newcommand{\Ham}{\mathcal{H}}
\newcommand{\Ord}{\mathcal{O}}

\newcommand{\Chg}{\mathcal{Q}}

\newcommand{\FS}{\mathcal{V}}

\newcommand{\met}{g_{\alpha\beta}}

\newcommand{\mmet}{\eta_{\alpha\beta}}

\newcommand{\grav}{h_{\alpha\beta}}
\newcommand{\igrav}{h^{\alpha\beta}}
\newcommand{\Hab}{H_{\alpha\beta}}
\newcommand{\iHab}{H^{\alpha\beta}}
\newcommand{\hs}[1]{h_{#1}{}^{#1}}
\newcommand{\Hs}[1]{H_{#1}{}^{#1}}
\newcommand{\gdet}{\sqrt{-g}}

\newcommand{\expec}[1]{\left\langle #1 \right\rangle}
\newcommand{\trans}{\enskip\rightarrow\enskip}

\newcommand{\qwhere}{\quad\text{where}\quad}

\newcommand{\qcom}{\,,\quad}

\newcommand{\nreturn}{\nonumber \\[0.3em]}
\newcommand{\return}{\\[0.3em]}

\newcommand{\bc}{\begin{color}{red}}
\newcommand{\ec}{\end{color}}

\newcommand{\Com}[2]{[#1,#2]}

\newcommand{\bigCom}[2]{\big[#1,#2\big]}
\newcommand{\BigPB}[2]{\Big\{#1,#2\Big\}}
\newcommand{\BigCom}[2]{\Big[#1,#2\Big]}
\newcommand{\sml}[1]{\mbox{\large $#1$}}

\newcommand{\osc}[2]{\hat{#1}}
\newcommand{\oscd}[2]{\hat{#1}^\dagger}
\newcommand{\aop}[2]{\hat{a}_{#1,#2}}
\newcommand{\aopd}[2]{\hat{a}^\dagger_{#1,#2}}
\newcommand{\polt}[1]{\varepsilon_{{\!}_{\scriptstyle #1}}{}}

\newcommand{\half}{(\kern-.3pt1\kern-.8pt/\kern-.2pt2)}

\begin{document}

\title{	Spontaneous Conformal Symmetry Breaking and\\
		Quantum Quadratic Gravity}

\author{Jisuke \surname{Kubo}}
\email{kubo@mpi-hd.mpg.de}
\affiliation{Max-Planck-Institut f\"ur Kernphysik (MPIK), Saupfercheckweg 1, 69117 Heidelberg, Germany}
\affiliation{Department of Physics, University of Toyama, 3190 Gofuku, Toyama 930-8555, Japan}

\author{Jeffrey \surname{Kuntz}}
\email{jkuntz@mpi-hd.mpg.de}
\affiliation{Max-Planck-Institut f\"ur Kernphysik (MPIK), Saupfercheckweg 1, 69117 Heidelberg, Germany}

\date{\today}

\begin{abstract}
We investigate several quantum phenomena related to quadratic gravity after rewriting the general fourth-order action in a more convenient form that is second-order in derivatives and produces only first-class constraints in phase space. We find that a Higgs mechanism may occur in the conformally invariant subset of the general quadratic action if the theory is conformally coupled to a scalar field that acquires a non-zero vacuum expectation value and spontaneously breaks the conformal symmetry. Then, in the broken phase, the originally massless spin-2 ghost may absorb both the scalar and vector fields to become massive. We also perform a BRST quantization of second-order quadratic gravity in the covariant operator formalism and discuss conditions under which unitarity of the full interacting quantum theory may be established.
\end{abstract}


\maketitle

\clearpage

\section{Introduction}

Einstein's theory of General Relativity (GR) has famously passed every experimental test thrown at it, despite its apparent failure to represent an acceptable theory of quantum gravity. This failure is primarily due to the theory's massive coupling constant, a feature that renders it inherently power counting non-renormalizable after quantization. Naturally, it has long been the dream of theorists to establish an extension to GR that represents a predictive quantum theory that reproduces the predictions of GR in the classical limit, at least to the level of modern experiments. The most natural way to extend GR is to simply include second powers of curvature tensors in the action at the classical level, since these kind of terms inevitably appear anyway as quantum corrections to GR \cite{tHooft1974}. The most general of these actions, which contains the three independent squares of the Riemann tensor, is known as quadratic gravity (QG) \cite{Salvio2018}. 

QG passes the first test of a satisfactory quantum theory of gravity as it may be quantized in terms of a dimensionless coupling constant and is indeed renormalizable, as was shown by Stelle \cite{Stelle1977} (see also \cite{Buchbinder,Elizalde1995a,Elizalde1996} for studies of the renormalization group in QG). However, despite this important feature, there are complications that come from including quadratic powers of curvature tensors, namely, they generically lead to the presence of additional gravitational degrees of freedom (DOFs) as compared to GR. This is a consequence of the fact that the curvature-squared terms are necessarily fourth-order in derivatives. There are various ways to deal with the ``hidden" DOFs in QG, but for the purposes of this paper, we find it beneficial to expose them from the start via the introduction of auxiliary fields that return the original action after being integrated out. This kind of trick is common practice, but it alone is not enough to see all the propagating DOFs at the level of the action. After introducing auxiliary fields, we will also introduce Stückelberg fields (with associated gauge symmetries) in the style of \cite{Kugo2014}, a procedure that is equivalent to the exchange of second-class for first-class constraints in the Hamiltonian picture. The auxiliary and Stückelberg tricks paired together will leave us with an action where a full separation of the gravitational DOFs is clearly manifest. Construction of this second-order (first-class) formalism will be the subject of Section \ref{sec:secondorder}.

One of the principal benefits of the second-order formulation of QG that we will present is that it allows for a straightforward identification of a Higgs mechanism in conformal gravity. Conformal gravity is a particular case of quadratic gravity that is invariant under the usual local diffeomorphisms as well local conformal (and special conformal) transformations. This theory was put forth by Weyl as an alternative to GR shortly after its introduction and it remains intriguing to many theorists as it represents a proper gauge theory of the conformal group \cite{Kaku1977} and because it hints towards asymptotically free UV completions of the Standard Model and gravity when coupled to matter \cite{Buchbinder1989,tHooft2011,tHooft2015,Jizba,Salvio2018b}. However, our universe is obviously not conformally invariant, so if conformal gravity were to serve as a UV-complete theory of gravity, it is natural to assume that some kind of spontaneous symmetry breaking (SSB) mechanism must exist that leaves the low-energy theory invariant only under diffeomorphisms. In Section \ref{sec:Higgs}, we will demonstrate that exactly such a process can occur if the action for conformal gravity is conformally coupled to a scalar field that acquires a non-zero vacuum expectation value (VEV). Using our second-order formulation, it becomes apparent that below the scale of symmetry breaking, the massless spin-2 ghost propagated by the bare conformal gravity action can swallow both the scalar and our Stückelberg vector to become massive in a kind of  ``double Higgs mechanism". The resulting action may then be interpreted in a unitary gauge that manifests as the sum of the Einstein-Hilbert action and a ghostly Fierz-Pauli action describing massive gravity, even at full non-linear order.

Another useful feature of the second-order formulation shown here is that it allows for a straightforward yet rigorous quantization of QG. In Section \ref{sec:quant} we will use the methods of Kugo, Ojima, and Nakanishi \cite{Kugo1978-2,Nakanishi1990,Kugo,Kugo2014} to establish a covariant operator quantization of our theory under the BRST formalism that describes all fields in terms of simple-poles in the asymptotic limit. We apply this approach to the full general QG action since the results carry over easily to the particular case of spontaneously broken conformal gravity as well, with the only difference being that an additional gravitational scalar state (the scalaron) is present in the asymptotic spectrum in the QG case. BRST quantization of QG has been carried out in the past \cite{Kawasaki1981,Kawasaki1982,Kawasaki1983}, though our treatment will differ in significant ways, principally in our application of the covariant operator formalism, which paired with our second-order formulation, makes the Kugo-Ojima quartet mechanism \cite{Kugo1979b} and the subsequent identification of all physical asymptotic states readily apparent.

The biggest hurdle standing in the way of QG being considered an acceptable theory of quantum gravity is the ghost problem i.e.\ the risk of unitarity violation due to the presence of a massive spin-2 ghost, which is itself a quantum realization of the classical Ostrogradsky instability \cite{Ostrogradsky1850,Woodard2015}. Many interesting attempts have been made over the years to address this problem and show unitarity in QG, with some of the most notable examples being \cite{Boulware1983,Salvio2016,Anselmi2018,Bender2008,Mannheim2018,Salvio2019a,Salvio2021,Donoghue2021}. In Section \ref{sec:unitarity} of this work, we present another take on this issue and argue that the ghost problem only becomes relevant for QG at energies nearing the mass of the spin-2 ghost, which is itself only roughly a few orders of magnitude from the Planck scale. We will impose  kinematical conditions that allow us to identify a subspace of the whole Fock space of physical states that is unitary, where we assume for simplicity that the spin-2 ghost is stable. This notion of ``conditional unitarity" is based on the fact that massive ghosts cannot exist as asymptotic states at energies below their mass, meaning that the S-matrix for this low-energy subspace contains no spin-2 ghost states, despite the fact that the ghosts may be excited virtually.

\section{Classical quadratic gravity at second order} \label{sec:secondorder}

\subsection{Second order formulation from auxiliary fields}

The most general action describing quadratic gravity is given by a sum of the Einstein-Hilbert action and the three independent squares of the Riemann tensor\footnote{We use the metric signature $(-1,1,1,1)$ and the Riemann tensor sign $\tensor{R}{_\alpha_\beta_\gamma^\delta}=-\partial_\alpha\tensor{\Gamma}{^\delta_\beta_\gamma}+\cdots$.}:
\begin{align} 
S = \int\dd^4x\gdet\bigg[\frac{M_\text{pl}^2}{2}R + aR_{\alpha\beta\gamma\delta}R^{\alpha\beta\gamma\delta} + bR_{\alpha\beta}R^{\alpha\beta} + cR^2\bigg] \,,
\end{align}
where $M_\text{pl}=(8\pi G)^{-1}$ is the reduced Planck mass, and $a$, $b$, and $c$ are arbitrary dimensionless constants. This action may be simplified by eliminating the Riemann square using the Gauß-Bonnet invariant
\begin{align}
\mathcal{G} = R_{\alpha\beta\gamma\delta}R^{\alpha\beta\gamma\delta} - 4R_{\alpha\beta}R^{\alpha\beta} + R^2 \,,
\end{align}
which is a total derivative i.e.\ a boundary term that may be set to zero at the level of the action \cite{Alvarez-Gaume2016}. With this we may eliminate one arbitrary constant and, after redefining the other two in terms of the new constants $\alpha_g$ and $\beta$, we are left with the following form for our action.
\begin{align} \label{SQG}
S_\text{QG} = \int\dd^4x\gdet\bigg[\frac{M_\text{pl}^2}{2}R - \frac{1}{\alpha_g^2}\bigg(R_{\alpha\beta}R^{\alpha\beta} - \frac{1}{3}R^2 + \beta R^2\bigg)\bigg]
\end{align}
This specific paramaterization of the constants has been chosen so that $\alpha_g$ may play the role of a perturbation parameter when we linearize the action and so that $\beta$ parameterizes the non-conformally-invariant part of the quadratic action. The remaining quadratic part is equivalent to the action for Weyl's conformal gravity after dropping boundary terms, which may be shown using the identity 
\begin{align}
C_{\alpha\beta\gamma\delta}C^{\alpha\beta\gamma\delta} = 2R_{\alpha\beta}R^{\alpha\beta} - \frac{2}{3}R^2 + \mathcal{G} \,.
\end{align}

Some reshuffling of the degrees of freedom (DOF) in this action is in order since, as was demonstrated in \cite{Kubo2022}, covariant quantization of theories of gravity that are fourth order in derivatives is more easily carried out after replacing the fourth order terms in the action with a classically equivalent second order action. Accordingly, we introduce an auxiliary tensor field $H_{\alpha\beta}(x)$ and define the rank four ``metric'' $M_{\alpha\beta,\gamma\delta}$ and its inverse ${M^{-1}}^{\alpha\beta,\gamma\delta}$,
\begin{gather}
\begin{gathered}
M_{\alpha\beta,\gamma\delta} = \delta_{\alpha\beta\gamma\delta} - g_{\alpha\beta}g_{\gamma\delta} \qquad\qquad {M^{-1}}^{\alpha\beta,\gamma\delta} = \delta^{\alpha\beta\gamma\delta} - \frac13 g^{\alpha\beta}g^{\gamma\delta} \return
M_{\alpha\beta,\mu\nu}{M^{-1}}^{\mu\nu,\gamma\delta} = \delta_{\alpha\beta}^{\gamma\delta} \,,
\end{gathered}
\end{gather}
where
\begin{align} \label{deltaDef}
\delta_{\alpha\beta\gamma\delta} = g_{\alpha_{\scriptstyle (\gamma}}g_{{}_{\scriptstyle \delta)}\beta} \equiv \frac12\big(g_{\alpha\gamma}g_{\beta\delta} + g_{\alpha\delta}g_{\beta\gamma}\big)
\end{align}
is the rank-four identity matrix. Then, noting the identity
\begin{align}
R_{\alpha\beta}R^{\alpha\beta} - \frac13 R^2 = G_{\alpha\beta} {M^{-1}}^{\alpha\beta,\gamma\delta}G_{\gamma\delta}
\end{align}
where $G_{\alpha\beta}=R_{\alpha\beta}-\frac12 g_{\alpha\beta}R$ is the Einstein tensor, we add $\sqrt{-g}\,\frac14 H^{\alpha\beta}M_{\alpha\beta,\gamma\delta}H^{\gamma\delta}$
to the action (\ref{SQG}), which because of the equation of motion that follows ($H_{\alpha\beta}=0$), does not change the original action. This fact remains true if we make the shift $H \to  H + cM^{-1}G$ where $c$ is some constant, and as a result we can rewrite the Weyl tensor part of the action by adding a complete square term as
\begin{align}
-\frac{c^2}{4}GM^{-1}G + \frac14\big(MH + cG\big)^{\rm T}M^{-1}\big(MH + cG\big) = \frac{c}{2}G_{\alpha\beta}\iHab + \frac14 H^{\alpha\beta}M_{\alpha\beta,\gamma\delta}H^{\gamma\delta}
\end{align}
with $c=2\alpha_g^{-1}$. We may also pull a similar trick to reduce the $\beta R^2$ term to second order by introducing an auxiliary scalar field $\chi(x)$ and writing
\begin{align}
\frac{\beta}{\alpha_g^2}R^2 - \frac{1}{\beta}\left(\frac12\chi - \frac{\beta}{\alpha_g}R\right)^2 = \frac{1}{\alpha_g}R\chi - \frac{1}{4\beta}\chi^2 \,,
\end{align}
allowing us to replace the action (\ref{SQG}) with the auxiliary action
\begin{align} \label{Saux}
S_\text{aux} = \int\dd^4x\gdet\left[\frac{M_\text{pl}^2}{2}R + \frac{1}{\alpha_g}\Big(G_{\alpha\beta}\iHab + R\chi\Big) + \frac{1}{4}\Big(\Hab\iHab - \Hs{\alpha}\Hs{\beta}\Big) - \frac{1}{4\beta}\chi^2\right]
\end{align}
without changing any of the physics.

This and the original fourth order action (\ref{SQG}) are invariant under the local diffeomorphisms 
\begin{align} \label{gHdiff}
&g'_{\alpha\beta} = \met + \alpha_g\Lag_\xi\met &H'_{\alpha\beta} = H_{\alpha\beta} + \alpha_g\Lag_\xi H_{\alpha\beta} &&\chi' = \chi + \alpha_g\Lag_\xi\chi
\end{align}
where $\Lag_\xi$ is the Lie derivative in the direction of the arbitrary vector field $\xi^\alpha(x)$. These four symmetries mean that, in the Hamiltonian picture, $S_\text{aux}$ generates eight first-class constraints that allow us to eliminate sixteen of the forty two DOFs in phase space. It is well-known that quadratic gravity propagates eight independent DOFs in configuration space, namely a massless spin-2 graviton, a massive spin-2 ghost, and a massive scalar \cite{Alvarez-Gaume2016}. This then implies that there are ten second-class constraints generated by $S_\text{aux}$ according to Dirac's rule: $1/2(20+20+2-2*8-10)=8$.

\subsection{First-class constraints from Stückelberg fields}

The presence of second-class constraints can be inconvenient for covariant canonical quantization, however, as demonstrated in \cite{Kubo2022}, it is straight-forward to convert the second-class constraints to first class using a Stückelberg procedure. For quadratic gravity, this involves introducing a vector field $A_\alpha(x)$ and a scalar field $\pi(x)$ by applying the replacement
\begin{align} \label{Stuck}
\Hab \trans \Hab - \big(\nabla_\alpha A_\beta + \nabla_\beta A_\alpha\big) + \frac{2}{m}\nabla_\alpha\nabla_\beta\pi
\end{align}
to (\ref{Saux}), which leads to the final form of our action for second order quadratic gravity,
\begin{align} \label{SSOQG1}
S_\text{SOQG} = \int\dd^4x\gdet\bigg[&\frac{M_\text{pl}^2}{2}R + \frac{1}{\alpha_g}\Big(G_{\alpha\beta}\iHab + R\chi\Big) + \frac{1}{4}\Big(\Hab\iHab - \Hs{\alpha}\Hs{\beta}\Big) - \frac{1}{4\beta}\chi^2 \nreturn
&+ \frac{1}{4}F_{\alpha\beta}F^{\alpha\beta} + \Big(\nabla_\beta H_\alpha{}^\beta - \nabla_\alpha\Hs{\beta}\Big)\bigg(A^\alpha - \frac{1}{m}\nabla^\alpha\pi\bigg) \nreturn
&- R_{\alpha\beta}\bigg(A^\alpha - \frac{1}{m}\nabla^\alpha\pi\bigg)\bigg(A^\beta - \frac{1}{m}\nabla^\beta\pi\bigg) \bigg] \,,
\end{align}
where we have employed the contracted Bianchi identity $\nabla^\alpha G_{\alpha\beta}=0$, $m$ is an arbitrary mass scale to be identified later, and $F_{\alpha\beta} = \nabla_\alpha A_\beta - \nabla_\beta A_\alpha$ as usual. This new form for our action is still classically equivalent to the original fourth order action, a fact which may be confirmed by integrating $\Hab$ and $\chi$ from (\ref{Saux}) using the EOMs below.
\begin{gather}
\Hab = -\frac{2}{\alpha_g}\bigg(R_{\alpha\beta} - \frac{1}{6}\met R\bigg) + \nabla_\alpha A_\beta + \nabla_\beta A_\alpha - \frac{2}{m}\nabla_\alpha\nabla_\beta\pi \return
\chi = \frac{2\beta}{\alpha_g}R
\end{gather}

As part of the Stückelberg procedure performed above, our second order action has acquired additional gauge symmetries corresponding to each new field. $A_\alpha$ is associated with the vector symmetry
\begin{align} \label{vecsym}
\begin{aligned}
&g'_{\alpha\beta} = \met \qquad\qquad &&H'_{\alpha\beta} = \Hab + \nabla_\alpha\zeta_\beta + \nabla_\beta\zeta_\alpha \qquad\qquad &&\chi' = \chi \return
&A'_\alpha = A_\alpha + \zeta_\alpha &&\pi' = \pi \,,
\end{aligned}
\end{align}
while $\pi$ is associated with the scalar symmetry
\begin{align} \label{scalsym}
\begin{aligned}
&g'_{\alpha\beta} = \met \qquad\qquad\qquad &&H'_{\alpha\beta} = \Hab \qquad\qquad\qquad &&\chi' = \chi \return
&A'_\alpha = A_\alpha + \nabla_\alpha\sigma  &&\pi' = \pi + m\sigma
\end{aligned}
\end{align}
where $\zeta^\alpha(x)$ and $\sigma(x)$ are arbitrary vector and scalar fields respectively. Needless to say, our action is still diffeomorphism invariant after introducing the Stückelberg fields which transform with Lie derivatives in the same style as (\ref{gHdiff}). Thus, after the Stückelberg procedure our action contains twenty six fields and has nine gauge symmetries, meaning we can count $1/2(20+20+2+8+2-2*18)=8$ DOFs with no second class constraints, as desired. 

\section{The Higgs mechanism in conformal gravity} \label{sec:Higgs}

\subsection{Spontaneous breaking of conformal symmetry}

Before proceeding with a full BRST quantization of the classical theory presented in the last section, it is interesting to first look at an enlightening feature of the second order formalism presented there, namely that it allows us to identify how a kind of Higgs mechanism may occur with respect to conformal symmetry. We consider a more symmetric subset of the general theory in the last section by dropping the Einstein-Hilbert term and setting $\beta=0$ in the action (\ref{SQG}) to arrive at the following action describing Weyl's conformal gravity (after again dropping total derivatives)
\begin{align} \label{SCG}
S_\text{CG} = -\frac{1}{\alpha_g^2}\int\dd^4x\gdet\bigg(R_{\alpha\beta}R^{\alpha\beta} - \frac{1}{3}R^2\bigg) \,.
\end{align}
In addition to this purely gravitational action, we also consider the following action describing a real matter scalar $\phi(x)$ conformally coupled to gravity.
\begin{align} \label{Sphi}
S_\phi = \int\dd^4x\gdet\bigg(\frac{1}{2}\nabla_\alpha\phi\nabla^\alpha\phi + \frac{1}{12}\phi^2R\bigg)
\end{align}
Both of these actions are diffeomorphism invariant and invariant under the Weyl gauge transformations 
\begin{align} \label{Weyl}
&g'_{\alpha\beta} = e^{\alpha_g\omega}\met &&\phi' = e^{-\alpha_g\omega/2} \phi
\end{align}
where $\omega(x)$ is an arbitrary scalar.

It is known that conformal gravity (\ref{SCG}) propagates six independent massless DOFs: a spin-2 graviton, a spin-2 ghost, and a spin-1 vector \cite{Riegert1984,Kubo2022}, so we may anticipate seven total DOFs from the sum of the actions (\ref{SCG}) and (\ref{Sphi}). To arrive at a second-order first-class description of the theory with these DOFs exposed, we follow the procedures outlined in the last section, though for the case of conformal gravity, it is only necessary to introduce the auxiliary tensor field $\Hab$ and vector Stückelberg field $A_\alpha$ with the associated vector symmetry as in \cite{Kubo2022}. The resulting action is then equivalent to (\ref{SSOQG1}) after dropping the Einstein-Hilbert term and setting $\chi=\pi=0$.
\begin{align} \label{SSOCG}
S_\text{SOCG} = \int\dd^4x\gdet\bigg[&\frac{1}{\alpha_g}G_{\alpha\beta}\iHab + \frac{1}{4}\Big(\Hab\iHab - \Hs{\alpha}\Hs{\beta}\Big) + \frac{1}{4}F_{\alpha\beta}F^{\alpha\beta} \nreturn
&+ A_\alpha\Big(\nabla_\beta H^{\alpha\beta} - \nabla^\alpha\Hs{\beta}\Big)- R_{\alpha\beta}A^\alpha A^\beta\bigg]
\end{align}
The complete action of interest is then given by
\begin{align} \label{SSOCGphi}
S_\text{SOCG$\phi$} = S_\text{SOCG} + S_\phi \,,
\end{align}
which is invariant under standard diffeomorphisms as in (\ref{gHdiff}), the Stückelberg symmetry (\ref{vecsym}), and Weyl symmetry. Taking all these fields and gauge symmetries into account one finds $1/2(20+20+8+2-2*18)=7$ DOFs with no second class constraints, as expected.

There are no massive DOFs present in the action (\ref{SSOCGphi}) (a requirement for manifest conformal symmetry), however, similarly to the well-known spontaneous symmetry breaking that occurs in the Standard Model, force carriers may acquire a mass if the scalar coupled to them picks up a non-zero vacuum expectation value. Assuming that the same situation occurs with the scalar $\phi$, we may reparameterize it as
\begin{align} \label{VEV}
\phi = \frac{\mu}{\alpha_g} + \varphi
\end{align} 
where $\mu$ is a dimensionful constant that parameterizes the VEV $\expec{\phi}=\mu/\alpha_g$ and $\varphi(x)$ represents fluctuations around that minimum. In this scenario, the scalar part of the action becomes
\begin{align} \label{Sf}
S_\phi \overset{\text{SSB}}{\trans} S_\varphi = \int\dd^4x\gdet\bigg[\frac{1}{2}\nabla_\alpha\varphi\nabla^\alpha\varphi + \frac{1}{12}\bigg(\varphi^2 + \frac{2\mu}{\alpha_g}\varphi + \frac{\mu^2}{\alpha_g^2}\bigg)R\bigg]
\end{align}
where the Weyl gauge symmetry is maintained with the transformation rule
\begin{align}
&\varphi' = e^{-\alpha_g\omega/2}\bigg(\frac{\mu}{\alpha_g} + \varphi\bigg) - \frac{\mu}{\alpha_g}
\end{align}
which may be inferred from (\ref{Weyl}).
 
\subsection{Double Higgs mechanism}

After SSB, we see that an Einstein-Hilbert term (the last term in (\ref{Sf})) is generated, implying that one of the spin-2 states becomes massive \cite{Stelle1977}. Since massive spin-2 states possess five DOFs, it is clear that some kind of Higgs mechanism is in effect. To see this explicitly, we identify the quadratic part of the action after writing the metric as  perturbations around the flat space Minkowski metric $\mmet$,
\begin{align} \label{lingrav}
\met \trans \mmet + \alpha_g\grav \,,
\end{align}
where we have assumed that $\alpha_g$ is small so that it may serve as a perturbation parameter. After performing the linearization, all indices are to be contracted with the background metric $\mmet$, and the quadratic (free) part of the action (\ref{SSOCGphi}) is given by the following.
\begin{align} \label{S0}
S_0 = \int\dd^4x&\left[-\igrav\mathcal{E}_{\alpha\beta\gamma\delta}\bigg(\frac{\mu^2}{24}h^{\gamma\delta} - H^{\gamma\delta}\bigg) + \frac{1}{4}\Big(\Hab\iHab - \Hs{\alpha}\Hs{\beta}\Big) \right. \nreturn
&\;\,\left. + \frac{1}{4}F_{\alpha\beta}F^{\alpha\beta} + A_\alpha\Big(\partial_\beta\iHab - \partial^\alpha\Hs{\beta}\Big) - \frac{1}{2}\varphi\Box\varphi - \frac{\mu}{6}\varphi\Big(\mmet\Box - \partial_\alpha\partial_\beta\Big)\igrav\right]
\end{align}
Here, $\Box=\partial_\alpha\partial^\alpha$ is the d'Alembertian operator and $\mathcal{E}_{\alpha\beta\gamma\delta}$ is the flat space Lichnerowicz operator i.e.\ the kinetic operator of linearized General Relativity,
\begin{gather}
\left(\frac{\gdet}{2\kappa^2}\,R\right)\bigg\rvert_{\met \,\rightarrow\, \mmet + \kappa\grav} = -\frac{1}{4}\igrav\mathcal{E}_{\alpha\beta\gamma\delta}h^{\gamma\delta} + \Ord(\kappa) \return
\mathcal{E}_{\alpha\beta\gamma\delta} = -\frac12\Big(\big(\delta_{\alpha\beta\gamma\delta} - \eta_{\alpha\beta}\eta_{\gamma\delta}\big)\Box + \eta_{\alpha\beta}\partial_\gamma\partial_\delta + \eta_{\gamma\delta}\partial_\alpha\partial_\beta - 2\eta_{(\alpha_{\scriptstyle (\gamma}}\partial_{\beta)}\partial_{{}_{\scriptstyle \delta)}}\Big) \label{LichOp} \,,
\end{gather}
where $\kappa=M_\text{pl}^{-1}$ serves as a dimensionful perturbation parameter and $\delta_{\alpha\beta\gamma\delta}$ is defined as in (\ref{deltaDef}), with the Minkowski metric replacing the general metric.

The equation of motions (EOMs) that follow from the action (\ref{S0}) are 
\begin{align} 
&\mathcal{E}_{\alpha\beta\gamma\delta}h^{\gamma\delta} + \frac{1}{2}\Big(\Hab - \mmet\Hs{\gamma}\Big) - \frac{1}{2}\Big(\partial_\alpha A_\beta + \partial_\beta A_\alpha\Big) + \mmet\partial_\gamma A^\gamma = 0 \label{hEOM} \return
&\mathcal{E}_{\alpha\beta\gamma\delta}\Big(H^{\gamma\delta} - m^2h^{\gamma\delta}\Big) - \frac{m}{\sqrt{3}}\Big(\mmet\Box - \partial_\alpha\partial_\beta\Big)\varphi = 0 \label{HEOM} \return
&\Big(\mmet\Box - \partial_\alpha\partial_\beta\Big)A^\beta - \partial_\beta H_\alpha{}^\beta + \partial_\alpha\Hs{\beta} = 0 \label{AEOM} \return
&\Box\varphi  - \frac{m}{\sqrt{3}}\Big(\mmet\Box - \partial_\alpha\partial_\beta\Big)\igrav = 0 \,, \label{fEOM}
\end{align}
where we have identified the canonical mass scale $m=\mu/(2\sqrt{3})$. By looking at these EOMs, we find the following linear combination of the original fields that brings us to a scenario reminiscent of the massive Proca field parameterization after SSB in standard gauge theory.  
\begin{align} \label{PsiDef}
&\Psi_{\alpha\beta} = \frac{1}{m}\big(\Hab - \partial_\alpha A_\beta - \partial_\beta A_\alpha\big) + \frac{2}{\sqrt{3}\,m^2}\partial_\alpha\partial_\beta\varphi
\end{align}
With this definition, one may combine (\ref{hEOM}) and (\ref{HEOM}) to find the EOM for $\Psi_{\alpha\beta}$,
\begin{align} \label{PsiEOM}
\mathcal{E}_{\alpha\beta\gamma\delta}\Psi^{\gamma\delta} + \frac{m^2}{2}\big(\Psi_{\alpha\beta} - \mmet\Psi_\gamma{}^{\gamma}\big) = 0 \,,
\end{align}
which is precisely the EOM one finds in the Fierz-Pauli theory of massive spin-2 fields \cite{DeRham2014}. Though $\Psi_{\alpha\beta}$ is gauge-invariant, which is easily confirmed by transforming the right side of (\ref{PsiDef}), we may still constrain its degrees of freedom by deriving constraints from the EOM above. These indicate that our ``Proca'' field $\Psi_{\alpha\beta}$ obeys a simple massive Klein-Gordon EOM and possesses only five physical DOFs.
\begin{align} \label{PsiKG}
\partial_\beta\Psi_\alpha{}^\beta = 0 \,,\quad \Psi_\alpha{}^{\alpha} = 0 \qquad\Rightarrow\qquad \big(\Box - m^2\big)\Psi_{\alpha\beta} = 0
\end{align}

Thus, we see from (\ref{PsiEOM}) and (\ref{PsiKG}) that a double Higgs mechanism is in effect; the massless $H_{\alpha\beta}$ has eaten a massless vector \textit{and} a massless scalar, $A_\alpha$ and $\varphi$, to become the massive spin-2 field $\Psi_{\alpha\beta}$. In this version of the Higgs mechanism, $\varphi$ is the familiar would-be Nambu-Goldstone (NG) scalar while $A_\alpha$ is an artificially introduced NG vector. Naturally, the total number of DOFs are preserved in this process since before SSB there are $2+2+2+1=7$ physical DOFs, while afterwards one finds $2+5=7$. This double Higgs mechanism can be seen even more explicitly at the level of the action if we identify the analogous Stückelberg-invariant redefinition of the massless DOFs. 
\begin{align} \label{psiDef}
\psi_{\alpha\beta} = m\grav - \Psi_{\alpha\beta} + \frac{1}{\sqrt{3}}\mmet\varphi
\end{align}
This definition paired with (\ref{PsiDef}) has the effect of canonicalizing both the massless and massive sectors i.e.\ diagonalizing the quadratic action\footnote{Similar diagonalizing redefinitions of the massless and massive spin-2 in QG may be seen in \cite{Hindawi1996,Salvio2019}.} (\ref{S0}), which becomes
\begin{align} \label{S0t}
S_0 = \int\dd^4x\left[-\frac12\Big(\psi^{\alpha\beta}\mathcal{E}_{\alpha\beta\gamma\delta}\psi^{\gamma\delta} - \Psi^{\alpha\beta}\mathcal{E}_{\alpha\beta\gamma\delta}\Psi^{\gamma\delta}\Big) + \frac{m^2}{4}\Big(\Psi_{\alpha\beta}\Psi^{\alpha\beta} - \Psi_\alpha{}^\alpha\Psi_\beta{}^\beta\Big)\right] \,.
\end{align}
Naturally, this action delivers the EOM (\ref{PsiEOM}) as well as the standard GR graviton EOM for $\psi_{\alpha\beta}$. We observe that the NG fields $\varphi$ and $A_\alpha$ have disappeared from (\ref{S0t}) completely and no physical Higgs field is present. Furthermore, the part of (\ref{S0t}) describing the massive spin-2 field $\Psi_{\alpha\beta}$ has the opposite sign for its kinetic term with respect to its massless counterpart, indicating its well-known role as a ghost.

\subsection{Unitary gauge} \label{sec:unigauge}

There is an additional analogy between the present theory and standard gauge theory that is worth commenting on, namely, there exists a ``unitary gauge'' where the NG bosons are transformed out of the full interacting action. This should not come as a surprise given that we have already seen that the action (\ref{S0t}) is nothing but the quadratic part of the full nonlinear action in said unitary gauge. 

To derive the full action in this gauge, we begin with the second order action $S_\text{SOCG$\phi$}$ given in (\ref{SSOCGphi}) and perform a Stückelberg vector transformation with the parameter $\zeta_\alpha= - A_\alpha$ yielding
\begin{align}
&H'_{\alpha\beta} = H_{\alpha\beta} - \nabla_\alpha A_\beta - \nabla_\beta A_\alpha &&A'_\alpha = A_\alpha - A_\alpha = 0 \,,
\end{align}
in order to remove the vector field from the action. Next, we perform a Weyl transformation (with $A_\alpha$ already suppressed) using the parameter $\omega=2\alpha_g^{-1}\ln(\alpha_g\phi/\mu)$ which gives
\begin{align} \label{phiWeyl}
\phi' = \bigg(\frac{\mu}{\alpha_g\phi}\bigg)\phi = \frac{\mu}{\alpha_g}
\end{align}
and eliminates the scalar field from the action; a possibility that is not present in conventional gauge theories. The final unitary gauge action is then given by
\begin{align} \label{SU}
S_\text{U} = \int\dd^4x\gdet\bigg[\frac{m^2}{\alpha^2_g}R +\frac{1}{\alpha_g}G_{\alpha\beta}\iHab + \frac{1}{4}\Big(\Hab\iHab - \Hs{\alpha}\Hs{\beta}\Big)\bigg] \,,
\end{align}
where we have dropped the primes on all of the fields and set $m=\mu/(2\sqrt{3})$.

Viewing the present theory in this gauge also allows us to separate the massless and massive degrees of freedom even at the non-linear level. To see this, we shift the general metric in order to Taylor expand the action according to
\begin{align}
\met \trans \met + a\Hab \,,
\end{align}
where $a$ is an arbitrary constant parameterizing the expansion. For the Einstein-Hilbert portion we have
\begin{gather}
S_\text{EH}[g] = \frac{m^2}{\alpha^2_g}\int\dd^4x\gdet R \,,
\end{gather}
\begin{align}
S_\text{EH}\big[g + aH\big] =\; &S_\text{EH}[g]  +  a\int\dd^4x\frac{\delta S_\text{EH}[g] }{\delta\met}\Hab \nreturn
&+ \frac{a^2}{2}\int\dd^4x\dd^4y\frac{\delta^2 S_\text{EH}[g] }{\delta\met\delta g_{\gamma\delta}}\Hab H_{\gamma\delta} + \Ord\big(H^3\big) \return
=\; & \frac{m^2}{\alpha^2_g}\int\dd^4x\gdet\left(R - aG_{\alpha\beta}\iHab - \frac{a^2}{2}\iHab E_{\alpha\beta\gamma\delta}H^{\gamma\delta}\right) + \Ord\big(H^3\big) \,,
\end{align}
where $E_{\alpha\beta\gamma\delta}$ is the full non-linear version of the flat space Lichnerowicz operator\footnote{We thank Taichiro Kugo for pointing out this identity.} (\ref{LichOp}),
\begin{align}
E_{\alpha\beta\gamma\delta} &= \frac{1}{\gdet}\frac{\delta^2 S_\text{EH}[g] }{\delta\met\delta g_{\gamma\delta}} \return
&= -\frac12\bigg(\big(\delta_{\alpha\beta\gamma\delta} - g_{\alpha\beta}g_{\gamma\delta}\big)\nabla_\mu\nabla^\mu + g_{\alpha\beta}\nabla_\gamma\nabla_\delta + g_{\gamma\delta}\nabla_\alpha\nabla_\beta - 2g_{(\alpha_{\scriptstyle (\gamma}}\nabla_{\beta)}\nabla_{{}_{\scriptstyle \delta)}} \nreturn
&\,\quad + 2C_{\alpha_{\scriptstyle (\gamma}\beta_{\scriptstyle \delta)}} - \frac23\bigg(\delta_{\alpha\beta\gamma\delta} - \frac14 g_{\alpha\beta}g_{\gamma\delta}\bigg)R\bigg) \,,
\end{align}
and $C_{\alpha\beta\gamma\delta}$ is the Weyl tensor.

The auxiliary conformal gravity portion of the action may be expanded in the same fashion yielding
\begin{gather}
S_\text{CG}[g] = \int\dd^4x\gdet\bigg[\frac{1}{\alpha_g}G_{\alpha\beta}\iHab + \frac{1}{4}\Big(\Hab\iHab - \Hs{\alpha}\Hs{\beta}\Big)\bigg] \,,
\end{gather}
\begin{align}
S_\text{CG}\big[g + aH\big] =\; &S_\text{CG}[g] + a\int\dd^4x\frac{\delta S_\text{CG}[g]}{\delta\met}\Hab + \Ord\big(H^3\big) \return
=\; & \int\dd^4x\gdet\bigg(\frac{1}{\alpha_g}\Big(G_{\alpha\beta}\iHab + a\iHab E_{\alpha\beta\gamma\delta}H^{\gamma\delta}\Big) \nreturn
&\hspace{5.7em} + \frac{1}{4}\Big(\Hab\iHab - \Hs{\alpha}\Hs{\beta}\Big)\bigg) + \Ord\big(H^3\big) \,.
\end{align}
The sum of these two actions is then given by 
\begin{align} \label{SUdiag}
S_\text{EH} + S_\text{CG} = \int\dd^4x\gdet\bigg(&\frac{m^2}{\alpha^2_g}R + \frac{1}{2m^2}\iHab E_{\alpha\beta\gamma\delta}H^{\gamma\delta} \nreturn
&+ \frac{1}{4}\Big(\Hab\iHab - \Hs{\alpha}\Hs{\beta}\Big)\bigg) + \Ord\big(H^3\big)
\end{align}
after setting $a=\alpha_g/m^2$ to cancel the mixed $G_{\alpha\beta}\iHab$ terms. We are thus left with a diagonal action that is equivalent to the full unitary gauge action (\ref{SU}) up to terms $\Ord(H^3)$ and contains only a standard E-H contribution for the metric paired with a ghost-like massive spin-2 contribution for $\Hab$.

Crucially, this diagonalized action has been computed at the full non-linear level which implies that fluctuations around any background metric have no kinetic mixing with $H_{\alpha\beta}$. To see this explicitly, we can expand the metric around a general background $\bar{g}_{\alpha\beta}$ as
\begin{align}
\met \trans \bar{g}_{\alpha\beta} + \alpha_g\grav \,,
\end{align}
which after also normalizing by setting $\grav=m\psi_{\alpha\beta}$ and $\Hab=m^{-1}\Psi_{\alpha\beta}$, yields
\begin{align} \label{SUdiaglin}
S_\text{EH} + S_\text{CG} = \int\dd^4x\sqrt{-\bar{g}}\bigg(&\frac{m^2}{\alpha^2_g}\bar{R} -\frac12\Big(\psi^{\alpha\beta}\bar{E}_{\alpha\beta\gamma\delta}\psi^{\gamma\delta} - \Psi^{\alpha\beta}\bar{E}_{\alpha\beta\gamma\delta}\Psi^{\gamma\delta}\Big) \nreturn
&+ \frac{m^2}{4}\Big(\Psi_{\alpha\beta}\Psi^{\alpha\beta} - \Psi_\alpha{}^\alpha\Psi_\beta{}^\beta\Big)\bigg) + \Ord(\alpha_g) \,,
\end{align}
where the bars indicate quantities evaluated on the background metric. Naturally, this action matches (\ref{S0t}) exactly if we select a flat background metric. We also note that one may simply take a metric perturbation that depends on the two independent tensor fields $\psi_{\alpha\beta}$ and $\Psi_{\alpha\beta}$ from the start as
\begin{align} \label{perthat}
g_{\alpha\beta} \trans \bar{g}_{\alpha\beta} + \frac{\alpha_g}{m}\big(\psi_{\alpha\beta} + \Psi_{\alpha\beta}\big) \,,
\end{align}
which applied to the action (\ref{SU}), returns (\ref{SUdiaglin}) exactly after setting $\Hab=m^{-1}\Psi_{\alpha\beta}$.

The fact that we are able to separate the mixing of the massless and massive spin-two fields at both the quadratic and full non-linear level implies that one may define an ordinary ``Einstein'' metric for theories of quadratic gravity that carries only the standard two massless graviton degrees of freedom, as opposed to the original metric that appears in the fourth order formulation of the theory that carries additional hidden DOFs. The physical content of the theory in terms of the diagonalized fields is transparent and the issue of unitarity can be treated in a clear way if the theory is expressed in terms of the Einstein metric. In this way, we may say that the system is in the ``unitary picture''. It is also important to point out that viewing quadratic gravity in this unitary picture is only possible in the presence of an explicit mass scale i.e.\ only after SSB in the case of conformal gravity. However, there are disadvantages to describing quadratic gravity in this picture as it describes a system in which the ghostly massive spin-two field is coupled to the ordinary (non-renormalizable) Einstein-Hilbert action. As such, it is clear that manifest power-counting renormalizability is lost; there will be no propagators that behave like $1/p^4$ at large momenta.

In contrast to this situation, the $h$--$h$ propagator in the original fields does behave like $1/p^4$ in the UV. This situation is seen explicitly in \cite{Kubo2022} where quantization of conformal gravity is carried out without the assumption of SSB i.e.\ where diagonalization of the fields is not possible. In this non-diagonalized ``renormalizable picture'', power-counting renormalizability is manifest as one may expect given the fact that the original fourth order theory is known to be renormalizable\footnote{The fact that the separation (\ref{perthat}) can be performed at the non-linear level, even though $\psi$ and $\Psi$ enter this expression only linearly, implies that $\Psi$ can act as a kind of Pauli-Villars regulator. This further implies that the original Weyl-squared term may also be seen as a regulator \cite{Pottel2020a}.} \cite{Stelle1977}. In any case, we will not discuss the renormalizability issue further as it is beyond the scope of the present work, and will instead focus on a unitary-esque picture in what follows, as quantization in this picture turns out to be quite straightforward.

\section{BRST quantization} \label{sec:quant}

Our next task is to establish a rigorous BRST quantization of quadratic gravity, for which we will employ similar procedures to \cite{Kugo1978-2} and \cite{Kugo2014}, where covariant BRST quantizations of GR and massive spin-2 Fierz-Pauli theory (in an $R_{\xi}$-style gauge) have been proposed. Our starting point will be the second-order first-class action (\ref{SSOQG1}), restated here for convenience,
\begin{align} \label{SSOQG2}
S_\text{SOQG} = \int\dd^4x\gdet\bigg[&\frac{m^2}{\alpha_g^2}R + \frac{1}{\alpha_g}\Big(G_{\alpha\beta}\iHab + R\chi\Big) + \frac{1}{4}\Big(\Hab\iHab - \Hs{\alpha}\Hs{\beta}\Big) - \frac{1}{4\beta}\chi^2 \nreturn
&+ \frac{1}{4}F_{\alpha\beta}F^{\alpha\beta} + \Big(\nabla_\beta H_\alpha{}^\beta - \nabla_\alpha\Hs{\beta}\Big)\bigg(A^\alpha - \frac{1}{m}\nabla^\alpha\pi\bigg) \nreturn
&- R_{\alpha\beta}\bigg(A^\alpha - \frac{1}{m}\nabla^\alpha\pi\bigg)\bigg(A^\beta - \frac{1}{m}\nabla^\beta\pi\bigg) \bigg] \,,
\end{align}
where we have identified $m=\alpha_gM_\text{pl}/\sqrt{2}$ in line with the results from the last section. It should be noted that the more general procedures described in the following sections may easily be transferred to the (conformal gravity + SSB scalar) case as well. In this case, if one makes the Weyl transformation (\ref{phiWeyl}) to absorb the scalar from the start, then the only difference between the resulting quantum theory and what we are about to present is an independent (non-ghost) scalar sector related to the $\chi$ terms which may simply be neglected if one wishes to consider a manifestly conformal theory at high energies.

\subsection{Gauge-fixing}

To begin the quantization process, we fix our gauge freedom using the BRST prescription \cite{Becchi1975,Becchi1976}\footnote{We recommend \cite{Kugo1979b,Nakanishi1990} for thorough treatments of BRST symmetry's role in quantizing gauge theories.}. We introduce sets of bosonic Nakanishi-Lautrup (NL) fields $B_a=\{b_\alpha(x),$ $B_\alpha(x),$ $B(x)\}$, as well as fermionic ghosts $C^a=\{c^\alpha(x),$ $C^\alpha(x),$ $C(x)\}$ and anti-ghosts $\bar{C}_a=\{\bar{c}_\alpha(x),$ $\bar{C}_\alpha(x),$ $\bar{C}(x)\}$, where the three fields in each of these sets correspond to the diffeomorphism, Stückelberg vector, and Stückelberg scalar symmetries respectively. It is important to note that the anti-ghosts are independent of the regular ghosts and not related by Hermitian conjugation. Terms involving these sets of fields are added to the classical Lagrangian (\ref{SSOCG}) in order to gauge fix the theory and establish BRST symmetry. BRST transformations are generated by the BRST charge operator $\Chg$ and their specific forms are fixed so that the BRST transformation is nilpotent ($\Chg^2=0$). The associated BRST algebra is graded in terms of a field's ``ghost number'', with the ghosts and anti-ghosts assigned ghost numbers of one and minus one respectively, while the classical fields and NL bosons carry a ghost number of zero. All physically relevant quantities, including the total action, are then restricted to be of ghost number zero.

Under BRST symmetry, the classical fields $\Phi_A=\{\met,\,\Hab,\,\chi,\,A_\alpha,\,\pi\}$ in (\ref{SSOQG2}) transform linearly as
\begin{align}
\Phi'_A = \Phi_A + \epsilon\delta\Phi_A
\end{align}
where $\epsilon$ is a constant anti-commuting and anti-Hermitian parameter of the BRST transformation and $\delta\Phi_A$ is given by the sum of the infinitesimal gauge transformations, with the transformation parameters replaced by the associated (canonically normalized) ghost fields.
\begin{align}
\begin{aligned} \label{nlBRST1} 
&\delta\met = \frac{\alpha_g}{m}\big(\nabla_\alpha c_\beta + \nabla_\beta c_\alpha\big) \return
&\delta\Hab = m\big(\nabla_\alpha C_\beta + \nabla_\beta C_\alpha\big) + \frac{\alpha_g}{m}\big(\nabla_\gamma\Hab + H_{\alpha\gamma}\nabla_\beta + H_{\beta\gamma}\nabla_\alpha\big)c^\gamma \return
&\delta\chi = \frac{\alpha_g}{m}\,c^\alpha\nabla_\alpha\chi \return
&\delta A_\alpha = mC_\alpha + \nabla_\alpha C + \frac{\alpha_g}{m}\big(\nabla_\beta A_{\alpha} + A_{\beta}\nabla_\alpha\big)c^\beta \return
&\delta\pi = mC + \frac{\alpha_g}{m}\,c^\alpha\nabla_\alpha\pi
\end{aligned}
\end{align}
The new BRST fields also transform under this symmetry as follows. 
\begin{align}
\begin{aligned} \label{nlBRST2} 
&\delta b_\alpha = 0 &&\delta B_\alpha = 0 &&\delta B = 0 \return
&\delta c^\alpha = \frac{\alpha_g}{m}\,c^\beta\partial_\beta c^\alpha \quad &&\delta C^\alpha = \frac{\alpha_g}{m}\big(c^\beta\partial_\beta C^\alpha + C^\beta\partial_\beta c^\alpha\big) \quad &&\delta C =\frac{\alpha_g}{m}\,c^\alpha\partial_\alpha C \return
&\delta \bar{c}_\alpha = ib_\alpha &&\delta \bar{C}_\alpha = iB_\alpha &&\delta \bar{C} = iB
\end{aligned}
\end{align}

We may proceed by selecting convenient gauge fixing conditions $G_a=0$ for each symmetry that generate the corresponding gauge-fixing and Faddeev-Popov ghost actions via BRST transformation according to
\begin{align} \label{gfaction}
S_\text{gf} + S_\text{FP} = -i\int\dd^4x\sqrt{-g}\,\delta\Big(\bar{C}^aG_a\Big) \,.
\end{align}
For diffeomorphism invariance we select the general condition
\begin{align} \label{nldiffcon}
G^{(\xi)}_\alpha = g^{\beta\gamma}\bigg(\partial_\gamma\tilde{g}_{\alpha\beta} - \frac{g_1}{2}\partial_\alpha\tilde{g}_{\beta\gamma}\bigg) + \frac12b_\alpha \,,
\end{align}
where
\begin{align}
\tilde{g}_{\alpha\beta} = \frac{m}{\alpha_g}\met - \frac{1}{m}\big(\Hab - \met\chi - \nabla_\alpha A_\beta - \nabla_\beta A_\alpha\big) - \frac{2}{m^2}\nabla_\alpha\nabla_\beta\pi \,,
\end{align}
is analogous to the Einstein metric discussed in section \ref{sec:unigauge} in the sense that it is Stückelberg symmetry invariant and only transforms under standard diffeomorphisms. In the same spirit, the Stückelberg vector symmetry is fixed with the condition
\begin{align} \label{nlvectorcon}
G^{(\zeta)}_\alpha = \frac{1}{m}\bigg(\nabla_\beta H_{\alpha}{}^\beta - \frac{g_2}{2}\nabla_\alpha\Hs{\beta} + \nabla_\alpha\chi\bigg) - mA_\alpha + \nabla_\alpha\pi -\frac{1}{2}B_\alpha \,,
\end{align}
which is diffeomorphism and scalar symmetry invariant. For the scalar symmetry\footnote{We note that $C$ and $\bar{C}$ would instead be FP ghosts corresponding to Weyl invariance if we would have quantized the conformal action (\ref{SSOCGphi}) using the method outlined here. In either case, $C$ and $\bar{C}$ are both propagating, as has also been observed in the unbroken version of second-order conformal gravity \cite{Kubo2022} and in the conformal theory without a Weyl-squared term \cite{Oda2022}, which should be contrasted with the discussions reviewed in \cite{Rachwa2022}.} we select the diffeomorphism and vector symmetry invariant condition
\begin{align} \label{nlscalarcon}
G^{(\sigma)} = \nabla_\alpha A^\alpha - \frac{g_3}{2}\Hs{\alpha} - \chi - m\pi - \frac{B}{2} \,.
\end{align}
The $g_i$ that appear in the conditions above are arbitrary constants that will allow us to investigate a few interesting gauge choices. The sum of the classical action (\ref{SSOQG2}) and the actions generated by applying (\ref{gfaction}) to each of the three conditions above gives us the gauge-fixed total action $S_\text{T}$.
\begin{align} \label{ST}
S_\text{T} &= S_\text{SOQG} -i\int\dd^4x\sqrt{-g}\,\delta\Big(\bar{c}^\alpha G^{(\xi)}_\alpha + \bar{C}^\alpha G^{(\zeta)}_\alpha + \bar{C} G^{(\sigma)}\Big) \nreturn
&= S_\text{SOQG} + S_{\text{gf}\xi} + S_{\text{gf}\zeta} + S_{\text{gf}\sigma} + S_{\text{FP}\xi} + S_{\text{FP}\zeta} + S_{\text{FP}\sigma}
\end{align}

\subsection{Free action and propagators}

With the full interacting total action established, our next task is to isolate the free part that is quadratic in the fields. We perturb the total action (\ref{ST}) around Minkowski space as in (\ref{lingrav}) and redefine the bare perturbation in terms of the massless ``Einstein'' graviton $\tilde{h}_{\alpha\beta}$ which eliminates all of its mixing with the massive sector as in (\ref{psiDef}).
\begin{align} \label{htDef}
&\tilde{h}_{\alpha\beta} = m\grav - \frac{1}{m}\big(\Hab - \met\chi - \partial_\alpha A_\beta - \partial_\beta A_\alpha\big) - \frac{2}{m^2}\partial_\alpha\partial_\beta\pi
\end{align}
It is also convenient to define normalized versions of the auxiliary fields as well as a new version of the Stückelberg scalar in order to diagonalize the scalar sector.
\begin{align} \label{HtDef}
&\tilde{H}_{\alpha\beta} = \frac{1}{m}\Hab &\tilde{\chi} = \frac{\sqrt{3}}{m}\chi &&\tilde{\pi} = \pi + \frac{1}{m}\chi
\end{align}
After linearizing (\ref{ST}), applying the redefinitions above, and dropping all of the $\Ord(\alpha_g)$ interaction terms, the free part of total action is given by the following.
\begin{align} \label{ST0}
S_0 = &\int\dd^4x\left[-\frac12\Big(\tilde{h}^{\alpha\beta}\mathcal{E}_{\alpha\beta\gamma\delta}\tilde{h}^{\gamma\delta} - \tilde{H}^{\alpha\beta}\mathcal{E}_{\alpha\beta\gamma\delta}\tilde{H}^{\gamma\delta}\Big) + \frac{m^2}{4}\Big(\tilde{H}_{\alpha\beta}\tilde{H}^{\alpha\beta} - \tilde{H}_\alpha{}^\alpha\tilde{H}_\beta{}^\beta\Big) \right. \nreturn
&+ \frac12\tilde{\chi}\Big(\Box - m_\beta^2\Big)\tilde{\chi} + \frac{1}{4}F_{\alpha\beta}F^{\alpha\beta} + mA_\alpha\Big(\partial_\beta\tilde{H}^{\alpha\beta} - \partial^\alpha\tilde{H}_\beta{}^\beta\Big) - \tilde{\pi}\Big(\mmet\Box - \partial_\alpha\partial_\beta\Big)\tilde{H}^{\alpha\beta} \nreturn
&+ b_\alpha\bigg(\partial_\beta\tilde{h}^{\alpha\beta} - \frac{g_1}{2}\partial^\alpha\tilde{h}_\beta{}^\beta + \frac12b^\alpha\bigg) + B_\alpha\bigg(\partial_\beta\tilde{H}^{\alpha\beta} - \frac{g_2}{2}\partial^\alpha\tilde{H}_\beta{}^\beta + \partial^\alpha\tilde{\pi} - mA^\alpha - \frac12B^\alpha\bigg) \nreturn
&+ B\bigg(\partial_\alpha A^\alpha - \frac{g_3m}{2}\tilde{H}_\alpha{}^\alpha - m\tilde{\pi} - \frac12B\bigg) + i\bigg(\bar{c}^\alpha\Big(\mmet\Box + (1 - g_1)\partial_\alpha\partial_\beta\Big)c^\beta \nreturn
&+\bar{C}^\alpha\Big(\mmet\big(\Box - m^2\big) + (1 - g_2)\partial_\alpha\partial_\beta\Big)C^\beta + \bar{C}\Big(\big(\Box - m^2\big)C + m(1-g_3)\partial_\alpha C^\alpha\Big)\bigg)\bigg]
\end{align}
Here, we have identified $m_\beta^2=m^2/(6\beta)$ as the canonical mass squared of the scalar $\tilde{\chi}$. It is interesting to note that in this parameterization, this scalar sector has been completely separated from the rest of the action. Since $\tilde{\chi}$ is also gauge invariant, it appears as nothing more than a basic massive scalar field, for which the quantization process is practically trivial compared to the rest of the fields in theory.

Moving forward, we note two specific choices of gauge fixing parameters that are worth considering. The first is given by
\begin{align} \label{unigauge}
g_1 = 1 \qquad\qquad g_2 = 2 \qquad\qquad g_3 = 0
\end{align}
which essentially corresponds to the unitary gauge presented in the last section. With this choice of parameters, one may make a further redefinition of $\tilde{H}_{\alpha\beta}$ in terms of the gauge-invariant tensor field
\begin{align} \label{UDef}
U_{\alpha\beta} = \tilde{H}_{\alpha\beta} - \frac{1}{m}\big(\partial_\alpha A_\beta + \partial_\beta A_\alpha\big) - \frac{1}{m^2}\big(\partial_\alpha B_\beta + \partial_\beta B_\alpha - 2\partial_\alpha\partial_\beta\tilde{\pi}\big) \,,
\end{align}
which eliminates the Stückelberg fields from the classical part of the action and removes the massive spin-2 field $U_{\alpha\beta}$ from the gauge conditions, leaving an independent massive spin-2 Fierz-Pauli action behind as in (\ref{SU}), with the resulting EOM and constraints as in (\ref{PsiKG}). The gauge choice (\ref{unigauge}) is enlightening in the sense that it allows one to see how the Stückelberg fields may be eaten by $\tilde{H}_{\alpha\beta}$, similarly to the SSB situation with the Stückelberg scalar $\pi$ now filling the role of the scalar $\varphi$. However, this gauge is not the most convenient for the upcoming calculations as the propagators become more complicated, making oscillator decomposition less straightforward.

The second interesting gauge choice, which is analogous to the Feynman gauge in standard gauge theory, remedies this issue and we will thus employ it for the remainder of this work.
\begin{align} \label{Feyngauge}
g_1 = 1 \qquad\qquad g_2 = 1 \qquad\qquad g_3 = 1
\end{align}
This choice of parameters yields a theory with only simple pole propagators, which we may see explicitly by looking at the full propagator matrix $\Omega^{-1}_{AB}(p)$. This matrix is the inverse of the Hessian matrix where $\Phi^A$ stands for the complete set of fields, including the NL bosons, ghosts, and anti-ghosts. It is given by
\begin{gather} \label{Hessian}
\Omega^{AB}(p) = i\int\dd^4x\frac{\delta^2 S_0}{\delta\Phi_A(x)\delta\Phi_B(y)}\,e^{-ip(x-y)} \return
\Omega^{-1}_{AB}(p) = -i\bra{0}T\Phi_A\Phi_B\ket{0} = \Array{
	  \Omega^{-1}_\text{boson} &  \mbox{\large $0$} \\
	 \mbox{\large $0$} & \Array{ \mbox{\footnotesize $0$} & \mbox{\footnotesize $ \Omega^{-1}_\text{ghost}$} \\ \mbox{\footnotesize $\Omega^{-1\,\dagger}_\text{ghost}$} & \mbox{\footnotesize $0$}}
	 }_{AB} \,,  \label{Prop}
\end{gather}
\\[-2\baselineskip]
\begin{align}
&\Omega^{-1}_\text{boson} = \nreturn
&\kbordermatrix{
	& \tilde{h}_{\gamma\delta} & \tilde{H}_{\gamma\delta} & \tilde{\chi} & A_\gamma & \tilde{\pi} & b_\gamma & B_\gamma & B \\
	\tilde{h}_{\alpha\beta} & \sml{\frac{-F_{\alpha\beta\gamma\delta}}{p^2}} & 0 & 0 & 0 & 0 & \sml{\frac{-i(\eta_{\alpha\gamma}p_\beta+\eta_{\beta\gamma}p_\alpha)}{p^2}} & 0 & 0 \\
	\tilde{H}_{\alpha\beta} &  & \sml{\frac{G_{\alpha\beta\gamma\delta}}{p^2+m^2}} & 0 & 0 & \sml{\frac{\eta_{\alpha\beta}}{3(p^2+m^2)}} & 0 & \sml{\frac{-i(\eta_{\alpha\gamma}p_\beta+\eta_{\beta\gamma}p_\alpha)}{p^2+m^2}} & 0 \\
	\tilde{\chi} &  &  & \sml{\frac{-1}{p^2+m_\beta^2}} & 0 & 0 & 0 & 0 & 0 \\
	A_\alpha &  &  &  & \sml{\frac{\eta_{\alpha\gamma}}{p^2+m^2}} & 0 & 0 & \sml{\frac{-m\eta_{\alpha\gamma}}{p^2+m^2}} & \sml{\frac{-ip_\alpha}{p^2+m^2}} \\
	\tilde{\pi} &  &  &  &  & \sml{\frac{1}{3(p^2+m^2)}} & 0 & 0 & \sml{\frac{-m}{p^2+m^2}} \\
	b_\alpha &  & \text{(h.c.)} &  &  &  & 0 & 0 & 0 \\
	B_\alpha &  &  &  &  &  &  & 0 & 0 \\
	B &  &  &  &  &  &  &  & 0
	} \\
&\Omega^{-1}_\text{ghost} = \kbordermatrix{
	& \bar{c}_\gamma & \bar{C}_\gamma & \bar{C} \\
	c_\alpha & \sml{\frac{-i\eta_{\alpha\gamma}}{p^2}} & 0 & 0 \\
	C_\alpha & 0 & \sml{\frac{-i\eta_{\alpha\gamma}}{p^2+m^2}} & 0 \\
	C & 0 & 0 & \sml{\frac{-i}{p^2+m^2}} 
	} \,,
\end{align}
where we have defined the shorthands
\begin{align}
&F_{\alpha\beta\gamma\delta} = 2\delta_{\alpha\beta\gamma\delta} - \eta_{\alpha\beta}\eta_{\gamma\delta} &&G_{\alpha\beta\gamma\delta} = 2\delta_{\alpha\beta\gamma\delta} - \frac23\eta_{\alpha\beta}\eta_{\gamma\delta} \,.
\end{align}
The utility of the Feynman gauge (\ref{Feyngauge}) is now clear; all of the individual propagators above contain only simple poles while maintaining nice behavior in the UV, mirroring what is expected from the covariant quantization of standard gauge theory \cite{Kugo1979b}, General Relativity \cite{Kugo1978-2}, and massive gravity \cite{Kugo2014} when similar gauges are employed. 

\subsection{Asymptotic fields}

We may proceed with the quantization process by appealing to the LSZ formalism \cite{Lehmann1955} in order to establish asymptotic solutions to the equations of motion of our system. In said formalism, one treats the fields in a theory $\Phi(x)$ as Heisenberg fields i.e.\ as quantum fields with time-independent state vectors, and makes the assumption that at times $t=x^0\to\pm\infty$, the $\Phi(x)$ behave as a free fields that satisfy the free equations of motion\footnote{Strictly speaking, the limit $x^0\to\pm\infty$ should be  a weak limit. We also ignore effects such as wave function renormalization here since they will not affect the essence of the work that follows.} \cite{Nakanishi1990}.
\begin{align}
\Phi(x) \to 
\left\{\begin{array}{c}
	\Phi^\text{in}(x) \,, \quad x^0 \to -\infty \\ 
	\Phi^\text{out}(x) \,, \quad x^0 \to +\infty
\end{array}\right.
\end{align}
The formalism dictates that each asymptotic field may be decomposed as a sum of products of oscillators and plane wave functions as
\begin{align} \label{PhiDecomp}
\Phi^{\text{as}}(x) = \sum_{\bm p}\bigg(\osc{\Phi}{f}^{\text{as}}({\bm p})f_{\bm p}(x,m) + \hat{\osc{\Phi}{g}}^{\text{as}}({\bm p})g_{\bm p}(x,m) + (\mbox{h.c.})\bigg) \qcom
\end{align}
where ${\bm p}$ stands for the three-dimensional spatial part of the four-momentum $p^\alpha$. Here, the plane wave functions $f_{\bm p}(x,m)$ and $g_{\bm p}(x,m)$ are solutions to the first and second order d'Alembert equations in the following sense.
\begin{align} \label{planewaves}
&\big(\Box - m^2\big)f_{\bm p}(x,m) = 0  &&\big(\Box - m^2\big)g_{\bm p}(x,m)  = f_{\bm p}(x,m)
\end{align}
The operator $\osc{\Phi}{f}^{\text{as}}({\bm p})$ in (\ref{PhiDecomp}) represents the fundamental simple-pole oscillator associated with the Heisenberg field $\Phi(x)$, where the superscript ``as'' stands for ``in'' or ``out'' depending on which limit is taken. These fundamental oscillators are products of annihilation(creation) operators and polarization tensors, which are non-trivial when $\Phi(x)$ is a field carrying space-time indices. The dipole oscillators (indicated with the double hat) are not independent DOFs, but rather functions of the fundamental oscillators that must be solved for using the EOMs. They are only non-zero if the associated field's propagator contains a dipole, which is not the case for the present theory in the gauge defined by (\ref{nldiffcon} -- \ref{nlscalarcon}, \ref{Feyngauge}).

The EOMs obtained from the total action (\ref{ST0}) for the bosons are found to be
\begin{align}
&\mathcal{E}_{\alpha\beta\gamma\delta}h^{\gamma\delta} + \partial_{(\alpha} b_{\beta)} - \frac12\mmet\partial_\gamma b^\gamma = 0 \return
&\mathcal{E}_{\alpha\beta\gamma\delta}H^{\gamma\delta} + \frac{m^2}{2}\Big(\Hab - \mmet\Hs{\gamma}\Big) - m\Big(\partial_{(\alpha} A_{\beta)} - \mmet\partial_\gamma A^\gamma\Big) \nreturn
&\qquad - \big(\mmet\Box - \partial_\alpha\partial_\beta\big)\pi - \partial_{(\alpha} B_{\beta)} + \frac12\mmet\partial_\gamma B^\gamma - \frac{m}{2}\mmet B = 0 \return
&\big(\Box - m_\beta^2\big)\chi = 0 \return
&\big(\mmet\Box - \partial_\alpha\partial_\beta\big)A^\beta - m\Big(\partial_\beta H_\alpha{}^\beta - \partial_\alpha\Hs{\beta}\Big) + mB_\alpha + \partial_\alpha B = 0  \return
&\big(\mmet\Box - \partial_\alpha\partial_\beta\big)\iHab + \partial_\alpha B^\alpha + mB = 0 \return
&\partial_\beta h_\alpha{}^\beta - \frac{1}{2}\partial_\alpha\hs{\beta} + b_\alpha = 0 \return
&\partial_\beta H_\alpha{}^\beta - \frac{1}{2}\partial_\alpha\Hs{\beta} - mA_\alpha + \partial_\alpha\pi - B_\alpha = 0 \return
&\Hs{\alpha} - \frac{2}{m}\big(\partial_\alpha A^\alpha - B\big) + 2\pi = 0 \,,
\end{align}
while the EOMs for the ghosts are given by 
\begin{align}
&\Box\,c_\alpha = 0 &&\Box\,\bar{c}_\alpha = 0 \return
&\big(\Box - m^2\big)C^\alpha = 0 &&\big(\Box - m^2\big)\bar{C}_\alpha= 0 \return
&\big(\Box - m^2\big)C = 0 &&\big(\Box - m^2\big)\bar{C} = 0 \,.
\end{align}
Note that here we have dropped all of the tilde designations from the fields for easier presentation, though one should keep in mind that we are only considering the canonical diagonal versions of our original fields as defined in (\ref{htDef}, \ref{HtDef}).

We know from the $p^{(-2n)}$ nature of the propagators (\ref{Prop}) that all of our fields contain only simple poles ($n=1$), indicating that they decompose in terms of simple oscillators as
\begin{align} \label{oscdecomp}
\begin{aligned}
&\grav(x) = \osc{h}{f}_{\alpha\beta}(\bm p)f_{\bm p}(x,0)  + \text{(h.c.)} \qquad\qquad&&\Hab(x) = \osc{H}{f}_{\alpha\beta}(\bm p)f_{\bm p}(x,m) + \text{(h.c.)} \return
&\chi(x) = \osc{\chi}{f}(\bm p)f_{\bm p}(x,m_\beta) + \text{(h.c.)} &&A_\alpha(x) = \osc{A}{f}_\alpha(\bm p)f_{\bm p}(x,m) + \text{(h.c.)} \return
&\pi(x) = \osc{\pi}{f}(\bm p)f_{\bm p}(x,m) + \text{(h.c.)} &&b_\alpha(x) = \osc{b}{f}_{\alpha}(\bm p)f_{\bm p}(x,0) + \text{(h.c.)} \return
&B_\alpha(x) = \osc{B}{f}_\alpha(\bm p)f_{\bm p}(x,m) + \text{(h.c.)} &&B(x) = \osc{B}{f}(\bm p)f_{\bm p}(x,m) + \text{(h.c.)} \return
&c^\alpha(x) = \osc{c}{f}^\alpha(\bm p)f_{\bm p}(x,m) + \text{(h.c.)} &&\bar{c}^\alpha(x) = \osc{\bar{c}}{f}^\alpha(\bm p)f_{\bm p}(x,m) + \text{(h.c.)} \return
&C^\alpha(x) = \osc{C}{f}^\alpha(\bm p)f_{\bm p}(x,m) + \text{(h.c.)} &&\bar{C}^\alpha(x) = \osc{\bar{C}}{f}^\alpha(\bm p)f_{\bm p}(x,m) + \text{(h.c.)} \return
&C(x) = \osc{C}{f}(\bm p)f_{\bm p}(x,m) + \text{(h.c.)} &&\bar{C}(x) = \osc{\bar{C}}{f}(\bm p)f_{\bm p}(x,m) + \text{(h.c.)}
\end{aligned}
\end{align}
where we have suppressed the sum over $\bm p$ as well as the ``as'' designations to avoid clutter.

Using (\ref{planewaves}), we see that all of the simple Klein-Gordon EOMs are satisfied by simply plugging in the decompositions (\ref{oscdecomp}), while the more complicated EOMs enforce the following additional conditions on the spin-2 oscillators.
\begin{align}
&p^\beta\osc{h}{f}_{\alpha\beta}(\bm p) = \frac{1}{2}p_\alpha\osc{h}{f}_\beta{}^\beta(\bm p) + i\osc{b}{f}_{\alpha}(\bm p) \label{transhf} \return
&p^\beta\osc{H}{f}_{\alpha\beta}(\bm p) = im\bigg(\frac{p_\alpha p_\beta}{m^2} - \mmet\bigg)\osc{A}{f}^\beta(\bm p) - 2p_\alpha\osc{\pi}{f}(\bm p) - i\osc{B}{f}_\alpha(\bm p) - \frac{1}{m}p_\alpha \osc{B}{f}(\bm p) \label{transHf} \return
&\osc{H}{f}_\alpha{}^\alpha(\bm p) = \frac{2}{m}\Big(ip^\alpha\osc{A}{f}_\alpha(\bm p) - \osc{B}{f}(\bm p)\Big) - 2\osc{\pi}{f}(\bm p) \label{traceHf}
\end{align}

In the continuum limit, (anti)commutators between each of the fundamental oscillators are given by the pole coefficient of the associated propagator entries in (\ref{Prop}) for the massless and massive fields respectively. The non-zero (anti)commutators are found to be
\begin{align}
&\BigCom{\osc{h}{f}_{\alpha\beta}(\bm p)}{\oscd{h}{f}_{\gamma\delta}(\bm q)} = \big(2\delta_{\alpha\beta\gamma\delta} - \eta_{\alpha\beta}\eta_{\gamma\delta}\big)\delta^3({\bm p} - {\bm q}) \return
&\BigCom{\osc{H}{f}_{\alpha\beta}(\bm p)}{\oscd{H}{f}_{\gamma\delta}(\bm q)} = \bigg(\!\!-\!2\delta_{\alpha\beta\gamma\delta} + \frac23\eta_{\alpha\beta}\eta_{\gamma\delta}\bigg)\delta^3({\bm p} - {\bm q}) \return
&\BigCom{\osc{\chi}{f}(\bm p)}{\oscd{\chi}{f}(\bm q)} = \delta^3({\bm p} - {\bm q}) \return
&\BigCom{\osc{A}{f}_\alpha(\bm p)}{\oscd{A}{f}_\beta(\bm q)} = -\mmet\delta^3({\bm p} - {\bm q}) \return
&\BigCom{\osc{\pi}{f}(\bm p)}{\oscd{\pi}{f}(\bm q)} = -\frac13\delta^3({\bm p} - {\bm q}) \return
&\BigCom{\osc{h}{f}_{\alpha\beta}(\bm p)}{\oscd{b}{f}_\gamma(\bm q)} = \BigCom{\osc{H}{f}_{\alpha\beta}(\bm p)}{\oscd{B}{f}_\gamma(\bm q)} = \big(ip_\alpha\eta_{\beta\gamma} +ip_\beta\eta_{\alpha\gamma}\big)\delta^3({\bm p} - {\bm q}) \return
&\BigCom{\osc{H}{f}_{\alpha\beta}(\bm p)}{\oscd{\pi}{f}(\bm q)} = -\frac13\mmet\delta^3({\bm p} - {\bm q}) \return
&\BigCom{\osc{A}{f}_\alpha(\bm p)}{\oscd{B}{f}_\beta(\bm q)} = m\mmet\delta^3({\bm p} - {\bm q}) \return
&\BigCom{\osc{A}{f}_\alpha(\bm p)}{\oscd{B}{f}(\bm q)} = ip_\alpha\delta^3({\bm p} - {\bm q}) \return
&\BigCom{\osc{\pi}{f}(\bm p)}{\oscd{B}{f}(\bm q)} = m\delta^3({\bm p} - {\bm q}) \return
&\BigPB{\osc{c}{f}_\alpha(\bm p)}{\oscd{\bar{c}}{f}_\beta(\bm q)} = \BigPB{\osc{C}{f}_\alpha(\bm p)}{\oscd{\bar{C}}{f}_\beta(\bm q)} = i\mmet\delta^3({\bm p} - {\bm q}) \return
&\BigPB{\osc{C}{f}(\bm p)}{\osc{\bar{C}}{f}(\bm q)} = i\delta^3({\bm p} - {\bm q}) \,.
\end{align}
With this relations in hand, all that remains is to appeal to the Kugo-Ojima quartet mechanism in order to identify which states are truly physical and which are unphysical remnants of the gauge freedom present in the original theory.

\subsection{Kugo-Ojima quartet mechanism}

In covariant BRST quantization, one may classify all of the quantum states in a theory into two distinct groups: BRST singlets, which are identified as physical states, and BRST quartets of unphysical states whose total contribution to any scattering amplitude always sums to zero. These quartets consist of pairs of ``parent'' states $\ket{\pi}$ and ``daughter'' states $\ket{\delta}$ that are related by BRST transformation as
\begin{align}
\ket{\delta_{g+1}} = \Chg\ket{\pi_g} \neq 0 \qcom
\end{align}
where the subscripts indicate FP ghost number. The precise way in which this cancellation of unphysical states occurs is known as the Kugo-Ojima quartet mechanism \cite{Kugo1978-1,Kugo1979a,Kugo1979}, the proof of which relies on showing that the following relationship between inner products of parents and daughters holds.
\begin{align} \label{qrel}
\braket{\pi_{-1}}{\delta_1} = \bra{\pi_{-1}}\Chg\ket{\pi_0} = \braket{\delta_0}{\pi_0} \neq 0
\end{align}
Demonstrating this equality explicitly is enough to guarantee that only physical transverse states will contribute to scattering amplitudes in a given theory. In order to show this for SOQG, we must simply reparameterize the states defined by the oscillators in the previous section in terms of BRST singlets and quartet participants.

\subsubsection{Massless states}

We begin with the massless spin-2 sector. For convenience, and without loss of generality, we restrict ourselves to a Lorentz frame defined by motion along the $z$-axis as defined by
\begin{align} \label{zframe}
p^\alpha = \big\{E,\,0,\,0,\,E\big\} \,,
\end{align}
recalling that all (anti)commutators derived in this frame are also valid in general \cite{Kugo1978-2}. In this basis, we may use the transverse oscillator equation (\ref{transhf}) to eliminate four of the ten components of $\osc{h}{f}_{\alpha\beta}$. From the remaining six independent components, we can identify that the two operators
\begin{align} \label{ahDef}
\aop{h}{+} = \frac{1}{2}\Big(\osc{h}{f}_{11} - \osc{h}{f}_{22}\Big) &&\aop{h}{\times} = \osc{h}{f}_{12}
\end{align}
are BRST singlets using the transformation rules\footnote{It is more convenient to express the BRST transformations of operators in terms of their (anti)commutators with the BRST charge operator $\Chg$ as we have done here. These two pictures are related by $\delta X=\Com{i\Chg}{X}_\mp$ where $\mp$ stands for commutator or anti-commutator as appropriate.} (\ref{nlBRST1}, \ref{nlBRST2}) combined with the redefintions (\ref{htDef}) and decompositions (\ref{oscdecomp}).
\begin{align}
\BigCom{\Chg}{\aop{h}{\lambda}} = 0 \quad\qwhere\quad \lambda = \{+,\,\times\}
\end{align}
The operators (\ref{ahDef}) have non-vanishing commutation relations only with themselves,
\begin{align} \label{ahComs}
\BigCom{\aop{h}{\lambda}(\bm p)}{\aopd{h}{\lambda'}(\bm q)} = \delta_{\lambda\lambda'}\delta^3({\bm p} - {\bm q}) \,,
\end{align}
and represent the physical transverse states contained in the massless graviton, which may be seen by rewriting the its simple pole oscillator as
\begin{align} \label{hfphys}
\osc{h}{f}_{\alpha\beta}(\bm p) = \polt{+}_{\alpha\beta}(\bm p)\aop{h}{+}(\bm p) + \polt{\times}_{\alpha\beta}(\bm p)\aop{h}{\times}(\bm p) + (\mbox{h.c.}) + \cdots \,,
\end{align}
where $\polt{\lambda}_{\alpha\beta}$ are transverse-traceless polarization tensors that, in the frame defined by (\ref{zframe}), may be written as the ``plus'' and ``cross'' forms familiar from General Relativity.
\renewcommand{\arraystretch}{.7}
\begin{align} \label{pmtens}
&\Big(\polt{+}_{\alpha\beta}\Big) = 
\begin{pmatrix}
0 & 0 & 0 & 0 \\
0 & 1 & 0 & 0 \\
0 & 0 & -1 & 0 \\
0 & 0 & 0 & 0
\end{pmatrix}
&&\Big(\polt{\times}_{\alpha\beta}\Big) = 
\begin{pmatrix}
0 & 0 & 0 & 0 \\
0 & 0 & 1 & 0 \\
0 & 1 & 0 & 0 \\
0 & 0 & 0 & 0
\end{pmatrix}
\end{align}
\renewcommand{\arraystretch}{1.0}

The ``$\cdots$'' in (\ref{hfphys}) represent contributions from the remaining four longitudinal components of $\osc{h}{f}_{\alpha\beta}$ which may be reparameterized in terms of the vector oscillator
\begin{align}
&\Big(\osc{\gamma}{f}_\alpha\Big) = \frac{i}{2E}\Array{
	-\osc{h}{f}_{00} \\
	2\osc{h}{f}_{13} \\
	2\osc{h}{f}_{23}\\
	\osc{h}{f}_{33}} \,.
\end{align}
These components are not BRST invariant and represent the $\ket{\pi_0}$ part of the quartets, while the roles of $\ket{\delta_0}$, $\ket{\delta_1}$, and $\ket{\pi_{-1}}$ are filled by $\osc{b}{f}_\alpha$, $\osc{c}{f}_\alpha$, and $\osc{\bar{c}}{f}_\alpha$ respectively.
\begin{align}
&\BigCom{\Chg}{\osc{\gamma}{f}_\alpha} = i\osc{c}{f}_{\alpha} &\BigCom{\Chg}{\osc{b}{f}_\alpha} = 0 &&\BigPB{\Chg}{\osc{c}{f}_\alpha} = 0 &&\BigPB{\Chg}{\osc{\bar{c}}{f}_{\alpha}} = \osc{b}{f}_\alpha
\end{align}
The commutation relations between these quartet participants are given by
\begin{align}
&\BigCom{\osc{\gamma}{f}_\alpha(\bm p)}{\oscd{b}{f}_\beta(\bm q)} = -\eta_{\alpha\beta}\delta^3(\bm p - \bm q) &&\BigPB{\osc{c}{f}_\alpha(\bm p)}{\oscd{\bar{c}}{f}_\beta(\bm q)} = i\eta_{\alpha\beta}\delta^3(\bm p - \bm q) \,,
\end{align}
indicating that the quartet mechanism functions as expected in the massless sector, as confirmed by the relation
\begin{align}
\mel{0}{\osc{b}{f}_\alpha(\bm p)\hat{\gamma}_\beta^\dagger(\bm q)}{0} = -i\mel{0}{\osc{\bar{c}}{f}_\alpha(\bm p)\oscd{c}{f}_\beta(\bm q)}{0} = -\mmet\delta^3(\bm{p - q}) \,.
\end{align}

\subsubsection{Massive states}

Identification of the quartets is also straightforward in the massive sector. Here it is more convenient to select the center of mass frame defined by
\begin{align} \label{cmframe}
p^\alpha = \big\{m,\,0,\,0,\,0\big\} \,,
\end{align}
which after employing the transverse and traceless constraints (\ref{transHf}, \ref{traceHf}), allows us to define the five operators
\begin{gather} \label{aHDef}
\begin{gathered}
\aop{H}{+} = \frac{1}{2}\Big(\osc{H}{f}_{11} - \osc{H}{f}_{22}\Big) \hspace{3em} \aop{H}{\times} = \osc{H}{f}_{12} \return
\aop{H}{1} = \osc{H}{f}_{13} \hspace{3em} \aop{H}{2} = \osc{H}{f}_{23} \hspace{3em} \aop{H}{3} = \frac{1}{2\sqrt{3}}\Big(\osc{H}{f}_{11} + \osc{H}{f}_{22} - 2\osc{H}{f}_{33}\Big) \,,
\end{gathered}
\end{gather}
in terms of the five independent components of $\osc{H}{f}_{\alpha\beta}$. Their commutation relations unavoidably come with a minus sign as compared to (\ref{ahComs}), indicating their ghost-like nature.
\begin{align} \label{ghostCom}
\BigCom{\aop{H}{\rho}(\bm p)}{\aopd{H}{\rho'}(\bm q)} = -\delta_{\rho\rho'}\delta^3({\bm p} - {\bm q})
\end{align}

Similarly to the massless case, (\ref{aHDef}) represents the BRST singlet components of $\osc{H}{f}_{\alpha\beta}$.
\begin{align}
\BigCom{\Chg}{\aop{H}{\rho}} = 0 \quad\qwhere\quad \rho = \{+,\,\times,\,1,\,2,\,3\}
\end{align}
These physical operators fit into the original oscillator as
\begin{align} \label{Hfphys}
\osc{H}{f}_{\alpha\beta}(\bm p) = \sum_\rho\Big(\polt{\rho}_{\alpha\beta}(\bm p)\aop{H}{\rho}(\bm p)\Big) + (\mbox{h.c.}) \,,
\end{align}
where there are naturally three physical longitudinal polarizations present in addition to the two transverse-traceless polarizations present in the massless case (\ref{pmtens}).
\renewcommand{\arraystretch}{.7}
\begin{align}
&\Big(\polt{1}_{\alpha\beta}\Big) = 
\begin{pmatrix}
0 & 0 & 0 & 0 \\
0 & 0 & 0 & 1 \\
0 & 0 & 0 & 0 \\
0 & 1 & 0 & 0
\end{pmatrix}
&\Big(\polt{2}_{\alpha\beta}\Big) = 
\begin{pmatrix}
0 & 0 & 0 & 0 \\
0 & 0 & 0 & 0 \\
0 & 0 & 0 & 1 \\
0 & 0 & 1 & 0
\end{pmatrix}
&&\Big(\polt{3}_{\alpha\beta}\Big) = 
\frac{1}{\sqrt{3}}\begin{pmatrix}
0 & 0 & 0 & 0 \\
0 & 1 & 0 & 0 \\
0 & 0 & 1 & 0 \\
0 & 0 & 0 & -2
\end{pmatrix}
\end{align}
\renewcommand{\arraystretch}{1.0}

All of the independent components of $\osc{H}{f}_{\alpha\beta}$ are accounted for in (\ref{Hfphys}), meaning that the remaining five components of our original classical fields may be assigned to the quartet participants
\begin{align}
&\Big(\osc{\Gamma}{f}_\alpha\Big) = -\frac{1}{m}\Array{
	\osc{A}{f}_{0} + i\osc{\pi}{f} \\
	\osc{A}{f}_{1} \\
	\osc{A}{f}_{2}\\
	\osc{A}{f}_{3}}
&&\osc{\Gamma}{f} = -\frac{1}{m}\osc{\pi}{f} \,.
\end{align}
These $\ket{\pi_0}$ operators and the other massive quartet partners transform in the expected fashions:
\begin{align}
&\BigCom{\Chg}{\osc{\Gamma}{f}_\alpha} = i\osc{C}{f}_{\alpha} &&\BigCom{\Chg}{\osc{B}{f}_\alpha} = 0 &&\BigPB{\Chg}{\osc{C}{f}_\alpha} = 0 &&\BigPB{\Chg}{\osc{\bar{C}}{f}_{\alpha}} = \osc{B}{f}_\alpha \return
&\BigCom{\Chg}{\osc{\Gamma}{f}} = i\osc{C}{f} &&\BigCom{\Chg}{\osc{B}{f}} = 0 &&\BigPB{\Chg}{\osc{C}{f}} = 0 &&\BigPB{\Chg}{\osc{\bar{C}}{f}} = \osc{B}{f} \,.
\end{align}
Their relevant non-vanishing commutation relations are found to be
\begin{align}
&\BigCom{\osc{\Gamma}{f}_\alpha(\bm p)}{\oscd{B}{f}_\beta(\bm q)} = -\eta_{\alpha\beta}\delta^3(\bm p - \bm q) \qquad &&\BigPB{\osc{C}{f}_\alpha(\bm p)}{\oscd{\bar{C}}{f}_\beta(\bm q)} = i\eta_{\alpha\beta}\delta^3(\bm p - \bm q) \return
&\BigCom{\osc{\Gamma}{f}(\bm p)}{\oscd{B}{f}(\bm q)} = -\delta^3(\bm p - \bm q) &&\BigPB{\osc{C}{f}(\bm p)}{\oscd{\bar{C}}{f}(\bm q)} = i\delta^3(\bm p - \bm q) \,,
\end{align}
which allows for another five realizations of the quartet mechanism in the massive sector.
\begin{gather}
\mel{0}{\osc{B}{f}_\alpha(\bm p)\oscd{\Gamma}{f}_\beta(\bm q)}{0} = -i\mel{0}{\osc{\bar{C}}{f}_\alpha(\bm p)\oscd{C}{f}_\beta(\bm q)}{0} = -\mmet\delta^3(\bm{p - q}) \return
\mel{0}{\osc{B}{f}(\bm p)\oscd{\Gamma}{f}(\bm q)}{0} = -i\mel{0}{\osc{\bar{C}}{f}(\bm p)\oscd{C}{f}(\bm q)}{0} = -\delta^3(\bm{p - q})
\end{gather}

Finally, as it is BRST invariant, $\osc{\chi}{f}$ represents an additional distinct physical state that does not participate in any of the quartets. It is also healthy (non-ghost-like) from a unitarity perspective, as is clear from its only non-vanishing commutation relation.
\begin{align}
&\BigCom{\Chg}{\osc{\chi}{f}} = 0 &&\BigCom{\osc{\chi}{f}(\bm p)}{\oscd{\chi}{f}(\bm q)} = \delta^3(\bm p - \bm q)
\end{align}
Thus, using BRST quantization and the quartet mechanism, we have identified the eight physical states present in quantum quadratic gravity. As expected, these are a total of three healthy states corresponding to the massless spin-2 $\aop{h}{\lambda}$ and massive spin-0 $\osc{\chi}{f}$ operators, and five ghost-like massive spin-2 states in $\aop{H}{\rho}$.

\subsection{Physical Hamiltonian operator}

With our physical states well-defined, we may ignore all of the unphysical quartet components and construct the physical (gauge-fixed) Hamiltonian operator $\Ham$ for our theory by solving the Heisenberg equation
\begin{align}
\bigCom{\Ham}{\phi(x)} = -i\partial_0\phi(x) \qcom
\end{align}
where $\phi(x)=\{h_{\alpha\beta}(x),\,H_{\alpha\beta}(x),\,\chi(x)\}$. After inserting the decompositions (\ref{oscdecomp}), the right side of this equation is easily determined using the relation
\begin{align}
&i\partial_0f_{\bm p}(x,m) = p^0f_{\bm p}(x,m) \,,
\end{align}
allowing us to infer the precise form of the Hamiltonian operator by looking at the Heisenberg equation for each $\phi(x)$.
\begin{align}
\Ham = \int\dd^3{\bm p}\sum_{\lambda,\rho}\Big(E_h\aopd{h}{\lambda}(\bm p)\aop{h}{\lambda}(\bm p) - E_H\aopd{H}{\rho}(\bm p)\aop{H}{\rho}(\bm p) + E_\chi\oscd{\chi}{f}(\bm p)\osc{\chi}{f}(\bm p)\Big)
\end{align}
This Hamiltonian is normal-ordered with respect to the vacuum as defined by
\begin{align}
\aop{h}{\lambda}(\bm p)\ket{0} = \aop{H}{\rho}(\bm p)\ket{0} = \osc{\chi}{f}(\bm p)\ket{0} = 0 \qcom
\end{align}
and it commutes with the state operators according to the relations
\begin{align}
&\BigCom{\Ham}{\aopd{h}{\lambda}(\bm p)} = E_h\aopd{h}{\lambda}(\bm p) &\BigCom{\Ham}{\aopd{H}{\rho}(\bm p)} = E_H\aopd{H}{\rho}(\bm p) &&\BigCom{\Ham}{\osc{\chi}{f}(\bm p)} = E_\chi\osc{\chi}{f}(\bm p) \,.
\end{align}
This makes it clear that each type of operator corresponds to a standard independent one-particle eigenstate with the eigenvalues $p^0$.

\section{Conditional unitarity in perturbative quadratic gravity} \label{sec:unitarity}
 
To demonstrate some notion of unitarity in the theory presented here, we follow the definitions of Kugo and Ojima \cite{Kugo1978-2}. This approach is again based on the LSZ formalism, which rests on the crucial assumptions that the Fock spaces spanned by the in and out states are both complete,
\begin{align}
\FS^{\text{in}} = \FS^{\text{out}} = \FS \,,
\end{align}
and that there exists an S-matrix operator $S$ that is pseudo-unitary ($S^\dag S=S S^\dag=\mathbbm{1}$) with elements defined by
\begin{align}
S_{\beta\alpha} = \braket{\beta;\text{out}}{\alpha;\text{in}} = \mel{\beta;\text{in}}{S}{\alpha;\text{in}} \,.
\end{align}
If $\FS$ is a positive-definite metric space, we may define unitarity in terms of the relation
\begin{align} \label{one}
1 &= \braket{\alpha;\text{in}}{\alpha;\text{in}} = \mel{\alpha;\text{in}}{S^\dag S}{\alpha;\text{in}} = \sum_n\mel{\alpha;\text{in}}{S^\dag}{n;\text{in}}\mel{n;\text{in}}{S}{\alpha;\text{in}} \nreturn
&= \sum_n\big|\mel{n;\text{in}}{S}{\alpha;\text{in}}\big|^2 \,,
\end{align}
where we have inserted the completeness relation $\mathbbm{1}=\sum_n\ket{n;\text{in}}\bra{n;\text{in}}$ between $S^\dag$ and $S$. A quantum theoretical probability interpretation follows from (\ref{one}), as it defines $|\mel{n;\text{in}}{S}{\alpha;\text{in}}|^2$ as the probability for the state transition $\alpha \to n$ to occur.

Generally speaking, in covariantly quantized gauge theories, the Fock space of physical states possesses a positive-definite metric even though $\FS$ as a whole is an indefinite-metric space. According to Kugo and Ojima \cite{Kugo1978-2}, this physical subspace $\FS_\text{phys}~(=\mbox{Ker}\, \Chg)$ is defined by 
\begin{align}
\Chg\ket{\text{phys}} = 0 \quad \forall \quad \ket{\text{phys}} \in \FS_\text{phys} \,.
\end{align}
Crucially, we have also have that $\FS_\text{phys}=S \FS_\text{phys}= S^\dag\FS_\text{phys}$ since $\FS_\text{phys}$ is invariant under time evolution as a result of $\Chg$ being a conserved charge. Furthermore, the quartet mechanism described in the previous section ensures that the zero-norm subspace ${\cal V}_0$ ($=\mbox{Im}\,\Chg$) of $\FS_\text{phys}$ is a BRST co-boundary. With this, the unitarity of $S$ on the quotient space $H_\text{phys}=\FS_\text{phys}/{\cal V}_0$ follows, provided that $H_\text{phys}$ is a \textit{positive-definite} metric space. Establishing unitarity in this context is equivalent to the statement that the quantum probability interpretation (\ref{one}) holds. 

The theory at hand represents a more complicated case, since in contrast to traditional gauge theories, the minus sign in (\ref{ghostCom}) indicates that $H_\text{phys}$ is a not positive-definite metric space. However, since this massive spin-2 ghost state is the only source of negative norm, we may define a positive-definite subspace by projecting the ghost states out from $H_\text{phys}$ kinematically. This is achieved by defining a basis for $H_\text{phys}$ that is spanned by the eigenstates $\ket{p_T,s}$ of the total four-momentum ${\cal P}^\mu_T$, where $s$ stands for other quantum numbers such as spin (or helicity). In this basis, we define the subspace $H_\text{phys}^{<}$ as the space spanned by the eigenstates $\ket{p_T,s}$ where
\begin{align} \label{kincon}
-p_T^2=-\eta_{\mu\nu}  \,p^\mu_T p^\nu_T < m^2 \,,
\end{align}
so that $H_\text{phys}^{<}$ contains no spin-2 ghost states. It is also important to note that $H_\text{phys}^{<}$ is Lorentz invariant simply because the whole $H_\text{phys}=\FS_\text{phys}/{\cal V}_0$ space is Lorentz invariant. This all implies that the S-matrix is unitary on the subspace $H_\text{phys}^{<}$, a feature that we will call ``conditional unitarity''. It should be noted that in the context of perturbation theory, one must replace the kinematic condition (\ref{kincon}) by
\begin{align}
p_T^2 + m^2 \gsim \Ord\big(m^2\big)\,,
\end{align}
because the spin-2 ghost propagator $\sim(p_T^2+m^2-i \epsilon)^{-1}$ should be sufficiently suppressed in order for perturbation theory to be valid. We remark that the notion of meta-stability such as that discussed in \cite{Salvio2019} does not influence this perturbative conditional unitarity because meta-stability is a non-perturbative effect and hence cannot be seen in perturbation theory.

So far we have assumed that the spin-2 ghost is stable, however, if it can decay as in \cite{Donoghue2019}, the spin-2 ghost states should be excluded from the asymptotic states anyway since in this case, the quotient space $H_\text{phys}$ becomes a positive-definite metric space without the need to impose any conditions that define an $H_\text{phys}^{<}$ subspace \cite{Veltman1963}. One might then conclude that the unitarity of $S$ must follow, however, this conclusion is premature. An important consequence of the unitarity of $S$ is the optical theorem, which states that the imaginary part of a forward scattering amplitude is given by the total cross section, where the associated Feynman diagrams are computed using the cutting rules defined by Cutkosky \cite{Cutkosky1960} and expanded upon by Veltman \cite{Veltman1963} to include unstable particles. Using a super-renormalizable scalar theory, Veltman has shown that only internal lines (propagators) of stable particles should be cut in order to satisfy unitarity while keeping the notions of renormalizability and causality intact. Veltman's treatment of unstable particles has been extended to the case of a complex pole mass of the propagator of an unstable particle in the context of perturbative gauge theories \cite{Denner2015}, however, the proof of this extension is given for ordinary healthy particles and it does not straightforwardly apply to the spin-2 ghost because the conditions imposed on ghostly propagators are different. In particular, it is assumed that the propagator of an unstable particle has a pole in the left upper-half of the complex $p^2$ plane in line with the conventional $i \epsilon$ prescription. However, this condition cannot hold for the spin-2 ghost propagator, a fact which has been demonstrated in the context of perturbation theory in \cite{Donoghue2019}\footnote{Similar situations present themselves in the Lee-Wick and Fakeon models \cite{Lee1969a,Anselmi2018}.}. 

In \cite{Donoghue2019}, an attempt has been made to show that the above-mentioned proof nevertheless holds for the spin-2 ghost with an exponentially suppressed violation of causality, though one should note that the proofs of \cite{Donoghue2019,Veltman1963,Denner2015} have been demonstrated only in scalar field theories. Our BRST-symmetry-based covariant quantization of QG, in which the massless graviton can be uniquely separated from the massive spin-2 ghost, will make a more rigorous treatment of the unitarity problem of quadratic gravity possible, though we will not go into further details here, leaving a thorough study of the issue in the covariant operator formalism for future work.

\section{Conclusion}

In this paper we have investigated several interesting phenomena related to quantum theories of quadratic gravity. To see these phenomena more transparently, we have rewritten the general fourth-order theory in an equivalent second-order form by introducing auxiliary fields, which paired with additional Stückelberg fields that render the phase-space constraints fully first-class, make the propagating DOFs apparent at the level of the action. In this second-order formalism, we have seen that a Higgs mechanism can occur in the conformally invariant subset of general quadratic gravity if the gravitational action is conformally coupled to a scalar field. If this scalar acquires a VEV and the conformal symmetry is spontaneously broken, the originally massless spin-2 ghost may eat the scalar as well as the Stückelberg vector, which in turn allows for the spontaneously broken theory to be interpreted in a ``unitary gauge". In this picture, the system describes massive (ghostly) spin-2 Fierz-Pauli theory coupled to the familiar Einstein-Hilbert action if the VEV of the scalar is assumed to be on the order of the Planck mass. We have also demonstrated that this separation of massless and massive spin-2 modes may be performed on arbitrary backgrounds and even at the full non-linear level.

The second-order formulation presented here also makes it possible for quadratic gravity to be straightforwardly quantized under the covariant operator formalism of Kugo, Ojima, and Nakanishi \cite{Kugo1978-2,Nakanishi1990,Kugo,Kugo2014}, which after appealing to the Kugo-Ojima quartet mechanism \cite{Kugo1979}, allows for an easy identification of the physically propagating DOFs in the quantum theory. In the Feynman-style gauge employed here, all asymptotic fields may described in terms of simple pole wave functions and we find that the physical Fock space is spanned by quantum states corresponding to a healthy massive scalar (the scalaron), a healthy massless spin-2 field (the traditional graviton), and a massive spin-2 ghost. This structure of the asymptotic spectrum appears as one should expect, while the explicit identification of an indefinite inner product metric on the physical spin-2 subspace of states (which is identical in either the spontaneously broken conformal or general quadratic case) allows for a new take on the ghost problem in quadratic quantum gravity.

In this work we do not attempt to provide a complete rigorous solution to the ghost problem, rather, we demonstrate the notion of ``conditional unitarity" by restricting the full spin-2 space of states with a simple kinematical condition that singles out a positive definite subspace. Put simply, we find that QG is a unitary theory at energies below the ghost mass $\lsim M_\text{pl}$. There is much more that can be said about this topic and so, in the interest of completeness, we have also included a brief appendix that contains more discussion on the ghost problem, conditional unitarity, and its relationship to the classical Ostrogradsky instability. Finally, it is worth mentioning once again that one may potentially make even stronger statements about unitarity in QG if the ghost is treated as an unstable particle, as in the interesting recent work of Donoghue and Menezes \cite{Donoghue2019}. An in-depth operator-based application of these ideas is certainly warranted, though we will leave this task to future work since this is a subtle and complicated topic with implications for many other areas of QFT that requires specific attention.

\clearpage

\textbf{Acknowledgments:} We wish to thank Taichiro Kugo and Manfred Lindner for many helpful discussions. J. Kubo is partially supported by the Grant-in-Aid for Scientific Research (C) from the Japan Society for Promotion of Science (Grant No.19K03844). We also note that the xAct suite of packages for Wolfram Mathematica were invaluable for performing the calculations shown here \cite{Martin-Garcia2008,Brizuela2009,Nutma2014a,Frob2020}.

\begin{appendix}

\section{Ostrogradsky instability and quantization} \label{sec:Ostro}

In this appendix we recapitulate the Ostrogradsky instability \cite{Ostrogradsky1850} and the associated problems of quantization by considering a concrete mechanical example. To understand the problem, it is important to discuss the problem at the classical level and the quantum level separately. It can quickly become confusing if one mixes up the problems of the two levels, as there are many (theoretical) phenomena in QFT that have no correspondence in classical theory.

We begin by considering a ``coordinate" in three spatial dimensions,  ${\bm x}(t)$, and assume its motion is described by the action
\begin{align} \label{LOst}
S_\text{Ost.} &= \int\dd t\left(\frac12\big(\dot{\bm x}(t)\big)^2 - \frac{1}{2\omega^2}\big(\ddot{\bm x}(t)\big)^2 - V\big(\bm x(t)\big)\right) \,,
\end{align}
which serves as a basic toy model of fourth-order quadratic gravity with the crucial feature being that, as it is a higher derivative theory, the Ostrogradsky instability must be present. The simple action (\ref{LOst}) belongs to a class of the Pais-Uhlenbeck model \cite{Pais1950} and its scalar field theory analogy has been considered in \cite{Salvio2019} at the classical level.

Though the Ostrogradsky instability can be discussed in terms of the fourth order action (\ref{LOst}), it is possible to transform the action into an equivalent form that allows us discuss the instability in a more transparent and concrete manner, similarly to our treatment of fourth-order quadratic gravity in the main text. To this end, we introduce an auxiliary ``coordinate"  ${\bm x}_2(t)$ and consider the action
\begin{align} \label{Laux}
S_\text{aux} &= \int \dd t \left(\frac12\big(\dot{\bm x}\big)^2 - \sqrt{m}\,\ddot{\bm x}\cdot {\bm x}_2 + \frac{m\omega^2}{2}\big({\bm x}_2\big)^2 - V({\bm x})\right) \,.
\end{align}
Inserting the equation of motion ${\bm x}_2=(1/\sqrt{m}\,\omega^2)\ddot{\bm x}$ into (\ref{LOst}), we see that (\ref{Laux}) is equivalent to (\ref{LOst}). Further, we may redefine ${\bm x}$ as
\begin{align} \label{x1}
{\bm x} = \sqrt{m}\big({\bm x}_1 - {\bm x}_2\big)
\end{align}
and insert this into (\ref{Laux}) to obtain the equivalent diagonalized action
\begin{align} \label{LU}
S_\text{U} = \int\dd t\left(\frac{m}{2}\big(\dot{\bm x_1}\big)^2 - \frac{m}{2}\big(\dot{\bm x_2}\big)^2 + \frac{m\omega^2}{2}\big({\bm x}_2\big)^2 - V\big(\sqrt{m}({\bm x}_1 - {\bm x}_2)\big)\right) \,,
\end{align}
where we have neglected total derivatives. This action with $V=0$ describes a system that consists of one free particle with the coordinate ${\bm x}_1$ and one harmonic oscillator with the coordinate ${\bm x}_2$, where the sign of the kinetic and mass terms for ${\bm x}_2$ is opposite to that of a ``healthy" coordinate. At this point, the analogy to the case of quadratic gravity is apparent: ${\bm x}_1$ corresponds to the massless graviton while ${\bm x}_2$ corresponds to the massive ghost spin-two field. This second order formalism is our starting point for investigating the ghost issue to which the conventional canonical Hamiltonian formalism can be applied. As we will see below, the real problem associated with the Ostrogradsky instability is not  the run away instability of the system, rather, it is that the classical system is pathological in the strict sense even if the run away instability is avoided.

\subsection{Classical level} 

The classical equations of motion derived from (\ref{LU}) are given by
\begin{align} \label{EOM12}
&m\frac{\dd^2{\bm x}_1}{\dd t^2} = -\nabla_1 V &&m\frac{\dd^2{\bm x}_2}{\dd t^2} = -m\omega^2 {\bm x}_2 + \nabla_2 V \,.
\end{align}
We see that due to the wrong sign for the kinetic term for ${\bm x}_2$, the force $\nabla_2 V$ has the wrong sign as compared to the force $-\nabla_1 V$, which is the origin of the Ostrogradsky instability. To see this more explicitly, let us assume that $V({\bm x})=V(|{\bm x}_1-{\bm x}_2 |)$ per the diagonalizing definition (\ref{x1}). Accordingly, we find
\begin{align} \label{force}
&-\!\nabla_1 V = \nabla_2 V = \frac{{\bm x}_2 - {\bm x}_1}{|{\bm x}_1-{\bm x}_2|}V' \,,
\end{align}
where $V'$ stands for $\dd V/\dd | {\bm x}_1-{\bm x}_2  |$. We see from (\ref{EOM12}) that if the force $-\nabla_1 V$ acting on ${\bm x}_1$ is a restoring force, then the force $\nabla_2 V$ acting on ${\bm x}_2$ must be an anti-restoring force. If this anti-restoring force is stronger than the other restoring force originating from the mass term $-m\omega^2 {\bm x}_2$, then the ghost particle will run away. However, we can also see that if $m\omega^2 |{\bm x}_2| > |\nabla_2 V| $ is satisfied for $|{\bm x}_2| > R$, where $R$ is a certain  finite constant, then the runaway of ${\bm x}_2$ can be avoided. It is also important to note that the exact form of the potential shown here is not such a crucial feature; it is just one of a few examples that show there are in fact many ways to avoid the run away instability \cite{Ilhan2013,Salvio2019,Deffayet2021}. 

The behavior just described is made even more apparent in the Hamiltonian formalism. The conventional definition of the canonical momenta yields ${\bm p}_1=m\dot{\bm x}_1$ and ${\bm p}_2 =-m\dot{\bm x}_2$, from which we obtain the classical Hamiltonian
\begin{align} \label{Hcl}
H = \frac{1}{2m}\big({\bm p}_1\big)^2 - \frac{1}{2m}\big({\bm p}_2\big)^2 - \frac{m\omega^2}{2}\big({\bm x}_2\big)^2 + V\big(|{\bm x}_1 - {\bm x}_2|\big) \,.
\end{align}
We can see that the total momentum ${\bm p}_1 + {\bm p}_2$ is conserved for $\omega=0$ because the potential is translationally invariant, however, the kinetic energy of the ghost particle is negative even though its motion can be stable as we have seen above. The total energy is also conserved, but due the negative kinetic energy of the ghost particle, strange situations can occur. 

Suppose, for instance, that the ghost particle sits at rest at the origin at $t=0$ i.e.\ ${\bm x}_2=\{0,0,0\}$ and $\dot{\bm x}_2=\{0,0,0\}$, and that the healthy particle runs parallel to the $x$-axis toward the ghost particle with some impact parameter ${\bm x}_1=\{-a,b,0\}$ and $\dot{\bm x}_1=\{v,0,0\}$. We may also assume that the potential is a gravity-like $1/r$ potential i.e.\ $V(|{\bm x}_1-{\bm x}_2|)=G |{\bm x}_1-{\bm x}_2|^{-1}$. The trajectory of the particles can be uniquely determined and the initial energy of the healthy particle is $E_\text{in} \simeq (1/2) m v^2$ if $a$ is large enough that the gravitational energy can be neglected, leaving $E_\text{in}$ as the approximate total energy. For $t>0$ the healthy particle approaches the ghost particle and scatters as a free particle, while the ghost particle leaves the origin and becomes a harmonic oscillator. After this scattering, the energy of the healthy particle, $E_\text{out}$, can become larger than the total energy because the energy of the ghost particle is negative and the total energy is conserved. This is precisely the famous Ostrogradsky ``instability". We emphasize once again that even in the presence of the ghost particle the motion can be stable (no run away) and deterministic, so that $E_\text{out}$ is fixed at a finite value for a given interaction, despite the fact that $E_\text{out}$ can be larger than $E_\text{in}$. Therefore, in the strict sense, the classical system is pathological even if the motion is stable. 

\subsection{Quantum level}

Given the classical Hamiltonian (\ref{Hcl}), it is straight forward to describe the corresponding quantum system in the Schrödinger picture using
\begin{align} \label{Schr}
 i\,\frac{\partial\Psi({\bm x}_1,{\bm x}_2,t)}{\partial t} = H\Psi({\bm x}_1,{\bm x}_2,t) \,,
\end{align}
where $\Psi({\bm x}_1,{\bm x}_2,t)$ is the wave function and
\begin{align} \label{Hschr}
H = -\frac{1}{2m}\big(\nabla_1\big)^2 + \frac{1}{2m}\big(\nabla_2\big)^2 - \frac{m\omega^2}{2}\big({\bm x}_2\big)^2 + V\big(|{\bm x}_1-{\bm x}_2|\big)
\end{align}
is the Hamilton operator for the quantum system. Although we do not prove it here, it is natural to assume that the quantum system in the Schrödinger picture approaches the classical system that we have described above in the classical limit. We emphasize that due to the nature of quantum mechanics, there is a tiny though non-zero probability that the energy of the outgoing healthy particle, $E_\text{out}$ in the scattering process described above, assumes an arbitrarily large value even if  the quantum system has no run away instability. Consequently, quantization in the Schr\"odinder picture fails to define a perfect quantum system, reflecting the pathology of the classical system in this picture.

We next consider the quantization of the system in the Heisenberg picture where, because of the wrong sign of the kinetic term of the ghost particle, it is possible to define a quantum system that has no classical correspondence. To this end, we first consider the free case $V=0$ and denote the momentum eigenstate of the healthy particle 
carrying the momentum ${\bm p}_1$ by $\ket{\bm p_1}$. For the harmonic oscillator (ghost state), we have the EOM $\ddot{\bm x}^\text{as}_2 +
 \omega^2{\bm x}^\text{as}_2=0$ with the solution
\begin{align}
x_{2,j}^\text{as}(t)  = \frac{1}{\sqrt{2m\omega}}\left(\hat{a}_j e^{-i\omega t} + \hat{a}_j^\dag  e^{i\omega t}\right) \,,
\end{align}
where $j=1,2,3$\, stands for the three spatial components of $\bm x_2$. From the canonical equal-time commutation relations $\left.\Com{x_{2,j}(t)}{p_{2,k}(t')}\right|_{t=t'}=i\delta_{jk}$, where $\bm p_2=-m\dot{\bm x}_2 $, we find
\begin{align} \label{a-comm}
\BigCom{\hat{a}_j}{\hat{a}_k^\dag} = -\delta_{jk} \,,
\end{align}
which has the wrong sign compared to the healthy harmonic oscillator. With this we can write the Hamiltonian operator $\hat{H}_0$ for the free system as
\begin{align} \label{H0}
\hat{H}_0 = -\frac{\omega}{2}\sum_{j=1,2,3} \left(\hat{a}_j\hat{a}_j^\dag + \hat{a}_j^\dag\hat{a}_j\right) = \omega\sum_{j=1,2,3} \left(-\hat{a}_j^\dag\hat{a}_j +\frac{1}{2}\right) = \omega\sum_{j=1,2,3} \left(-\hat{a}_j\hat{a}_j^\dag - \frac{1}{2}\right) \,.
\end{align}

So far we have encountered no ambiguities, however, some may arise due to the fact that there are two possibilities to define the vacuum (ground state):
\begin{align} \label{vacuum}
\hat{a}^\dag \ket{0}_- = 0 \qquad\text{or}\qquad \hat{a}\ket{0}_+ = 0 \,.
\end{align}
The second of these is the conventional definition for the healthy harmonic oscillator where it is the positivity of energy that dictates this choice. However, in the ghost case this is not mandatory because it is not necessary to satisfy positivity of the energy since the ghost's kinetic energy is negative already at the classical level. Excited states can also be defined in the standard fashion as  
\begin{align} 
\ket{n}_- = \frac{1}{\sqrt{n!}}\big(\hat{a}\big)^n\ket{0}_- \qquad\text{or}\qquad \ket{n}_+ = \frac{1}{\sqrt{n!}}\big(\hat{a}^\dag\big)^n \ket{0}_+ \,, 
\end{align}
which paired with the commutation relations (\ref{a-comm}) and the normalization conditions $\tensor[_-]{\braket{0}{0}}{_-}\!=\!\tensor[_+]{\braket{0}{0}}{_+}\!=1$, leads to to the products
\begin{align} \label{product}
\tensor[_-]{\braket{n'}{n}}{_-} = \delta_{n n'} \qquad \tensor[_+]{\braket{n'}{n}}{_+} = (-1)^n \delta_{n n'} \,.
\end{align}
It is therefore the second choice for the definition of the vacuum in (\ref{vacuum}) that forces us to deal with an indefinite metric. The difference between vacuum definition also appears in the energy eigenvalues
\begin{align} \label{H-eigen}
\hat{H}_0\ket{n}_- = -n\,\omega\ket{n}_- \qquad \hat{H}_0\ket{n}_+ = n\,\omega\ket{n}_- \,,
\end{align}
where we see that the the quantization based on $\ket{n}_-$ corresponds to that in the Schrödinger picture because the energy eigenvalues of the ghost are negative and no indefinite metric appears. Quantization based on the $\ket{n}_+$ on the other hand does not share this classical correspondence as the overall minus does not appear with the energy eigenvalues, but rather in the commutation relations i.e.\ with the indefinite metric.

Despite this lack of classical correspondence, we argue that the second choice for the vacuum, $\ket{n}_+$, is in fact the correct one. Consider the propagator
\begin{align}
\tensor[_\pm]{\mel{0}{T\hat{x}_j^\text{as}(t)\hat{x}_k^\text{as}(t')}{n}}{_\pm} = &\;\theta(t-t')\tensor[_\pm]{\mel{0}{\hat{x}_j^\text{as}(t)\hat{x}_k^\text{as}(t')}{n}}{_\pm} \nreturn
&+ \theta(t'-t)\tensor[_\pm]{\mel{0}{\hat{x}_j^\text{as}(t)\hat{x}_k^\text{as}(t')}{n}}{_\pm} \,,
\end{align}
which using the identity
\begin{align}
\theta (t) &=\frac{i}{2\pi}\int dE \,\frac{\,e^{-iE t}}{E+i\epsilon} \,,
\end{align}
may be written as
\begin{align} \label{propagator}
\tensor[_\pm]{\mel{0}{T\hat{x}_j^\text{as}(t)\hat{x}_k^\text{as}(t')}{n}}{_\pm} = -\delta_{jk}\frac{i}{2\pi m}\int\dd E\frac{e^{-iE(t - t')}}{E^2 - \omega^2\pm i\epsilon} \,.
\end{align}
The same propagator computed for the healthy particle takes the standard form and lacks the overall minus sign above, but there is an additional crucial difference between the two propagators, namely that they share the same pole structure for the $\ket{0}_+$ vacuum while the pole structure is opposite for the $\ket{0}_-$. It is opposite in the sense that for positive energy ($E=\omega$), the pole is located on the lower (upper) half complex plane for $\ket{0}_+$ ($\ket{0}_-$), while for negative energy it is located on the upper (lower) half complex plane for $\ket{0}_+$ ($\ket{0}_-$)\footnote{This fact should be regarded as a violation of causality for the $\ket{0}_-$ vacuum because the negative energy state propagates forward in time.}. Though $\epsilon$ is an infinitesimal parameter, it plays a very important roll in QFT; it is essential that all of the propagators in a renormalized Feynman diagram have the same $i\epsilon$ prescription in order to prove both the absolute convergence of the integration over the internal momenta with the Minkowski metric and the existence of the $\epsilon\to 0^+$ limit \cite{Zimmermann1968}  (see also \cite{Hepp1966,Appelquist1969}). Therefore, if healthy and ghost propagators are both present in a Feynman diagram, we have to obey the prescription for the healthy propagators (as suggested by Stelle \cite{Stelle1977} and see also Salvio \cite{Salvio2018}) which means that the correct choice of the vacuum is $\ket{0}_+$.

There is another reason that $\ket{0}_+$ is preferred. As we have seen, the energy eigenvalue of the ghost at the free level is positive and if we assume the existence of asymptotic states, this energy eigenvalue is the eigenvalue of the full Hamiltonian operator $\hat{H}$ in the Heisenberg picture. Additionally, since the S-matrix operator commutes with the full Hamiltonian operator, the asymptotic state $\ket{n}_+$ (which is an $\hat{H}$ eigenstate) remains after $S$ is applied. Thus, as an $\hat{H}$ eigenstate with the same eigenvalue, $S\ket{n}_+$ is also an $\hat{H}$ eigenstate. This feature is crucial for ensuring the conservation of energy, where neither "in" nor "out" states have a negative energy.

\subsection{Conditional unitarity for the quantum PU oscillator}

In the Heisenberg picture based on the vacuum $\ket{0}_+$, the Ostrogradsky instability appears as a violation of unitarity because of the indefinite metric structure (\ref{product}); since the norm is not positive definite, the probability interpretation of quantum theory fails. However, this does not mean that it is impossible to give any statement about the probability of a quantum process. It is in fact possible to give an exact statement if a certain condition is satisfied.

Before we address this statement in more detail, let us look at the classical system once again. We have seen that the classical motion can be stable if the restoring force acting on the ghost particle at the free level, $-m\omega^2 {\bm x}_2$, is stronger than the force from the potential $V(|{\bm x}_1-{\bm x}_2|)$. Indeed, if $V(r)$ approaches zero as $r\to\infty$, such a situation can easily be realized. Nevertheless, the negative energy of the ghost leads to an \textit{apparent} violation of energy conservation in the energy budget of the healthy particle. However, if $\omega$ is large, the healthy particle can excite the ghost harmonic oscillator only slightly in scattering processes implying an approximately elastic collision where approximate energy conservation with respect to the healthy particle can be realized.

With this observation in mind, we come back to the quantum system with the vacuum $\ket{0}_+$. In this system, the energy of the ghost harmonic oscillator is quantized and positive, and there is an energy gap between the vacuum and the first excited state $\omega$. If $\omega$ is large or if the energy of the incoming healthy particle is smaller than $\omega$, the ghost remains at the ground state. Of course, since this is a quantum process, the ghost can be virtually excited, but this has no influence on unitarity because only scalar products between on-shell states matter for unitarity. Thus, under the condition $E_\text{in}<\omega$, we can in fact make exact statements about the probability in a scattering processes of this system since $\ket{0}_+$ and all healthy states have positive norms, and the on-shell states can not contain excited ghost states. In this way the unitarity of the theory is exactly satisfied, provided that the scattering operator $S$ is a pseudo unitarity operator ($S^\dag\,S=1$) on the full space (including ghost states).
 
We note that the conclusion above is non-perturbative and may be extended to quadratic gravity, provided that we know about the stable classical solutions of its EOMs (which have been partially analyzed in \cite{Salvio2019}). On the other hand, in perturbation theory, the interactions are already assumed to be weak so that calculations at each order make sense. Therefore, within the framework of perturbation theory, where we assume the existence of the asymptotic fields, unitarity is exactly satisfied at each order if the kinematical constraint is satisfied; a notion referred to as ``conditional unitarity" in the main text. Additionally, if the excited ghost states are unstable, the kinematic condition may even be relaxed \cite{Donoghue2019}, however, a proper analysis of this fact in the present formalism is beyond the scope of this paper.

Finally, we recall that the choice of the vacuum $\ket{0}_+$ is dictated not just by a desire to show unitarity, but also by the proof of renormalization, a notion that has no classical correspondence. Similarly, in the $\ket{0}_+$ quantization, the appearance of an indefinite metric (negative norm) also has no classical correspondence. The Ostrogradsky instability manifests itself in the violation of unitarity, which is itself a consequence of the indefinite metric, however, as we have seen in the preceding mechanical example, the theory can be a perfect unitary theory for energies below the threshold of the ghost excitation even though the classical system is pathological in the strict sense.

\end{appendix}

\bibliographystyle{JHEP}
\bibliography{library}

\providecommand{\href}[2]{#2}\begingroup\raggedright\begin{thebibliography}{10}

\bibitem{tHooft1974}
G.~{'t Hooft} and M.~Veltman, {\it {One-loop divergencies in the theory of
  gravitation}},  {\em Ann. Inst. H. Poincare Phys. Theor. A} {\bf 20} (1974),
  no.~1 69--94.

\bibitem{Salvio2018}
A.~Salvio, {\it {Quadratic Gravity}},  {\em Frontiers in Physics} {\bf 6} (apr,
  2018) [\href{http://www.arxiv.org/abs/1804.09944}{{\tt 1804.09944}}].

\bibitem{Stelle1977}
K.~S. Stelle, {\it {Renormalization of higher-derivative quantum gravity}},
  {\em Physical Review D} {\bf 16} (aug, 1977) 953--969.

\bibitem{Buchbinder}
I.~L. Buchbinder, S.~D. Odintsov, and I.~L. Shapiro, {\em {Effective Action in
  Quantum Gravity}}.
\newblock IOP Publishing, sep, 1992.

\bibitem{Elizalde1995a}
E.~Elizalde, S.~D. Odintsov, and A.~Romeo, {\it {Improved effective potential
  in curved spacetime and quantum matter–higher derivative gravity theory}},
  {\em Physical Review D} {\bf 51} (feb, 1995) 1680--1691.

\bibitem{Elizalde1996}
E.~Elizalde, S.~D. Odintsov, and A.~Romeo, {\it {Renormalization group
  properties of higher-derivative quantum gravity with matter in 4 - $\epsilon$
  dimensions}},  {\em Nuclear Physics B} {\bf 462} (mar, 1996) 315--329.

\bibitem{Kugo2014}
T.~Kugo and N.~Ohta, {\it {Covariant approach to the no-ghost theorem in
  massive gravity}},  {\em Progress of Theoretical and Experimental Physics}
  {\bf 2014} (2014), no.~4 1--21.

\bibitem{Kaku1977}
M.~Kaku, P.~K. Townsend, and P.~{Van Nieuwenhuizen}, {\it {Gauge theory of the
  conformal and superconformal group}},  {\em Physics Letters B} {\bf 69}
  (1977), no.~3 304--308.

\bibitem{Buchbinder1989}
I.~L. Buchbinder, O.~K. Kalashnikov, I.~L. Shapiro, V.~B. Vologodsky, and J.~J.
  Wolfengaut, {\it {Asymptotic Freedom in the Conformal Quantum Gravity with
  Matter}},  {\em Fortschritte der Physik/Progress of Physics} {\bf 37} (1989),
  no.~3 207--223.

\bibitem{tHooft2011}
G.~{'t Hooft}, {\it {A Class of Elementary Particle Models Without Any
  Adjustable Real Parameters}},  {\em Foundations of Physics} {\bf 41} (2011),
  no.~12 1829--1856, [\href{http://www.arxiv.org/abs/1104.4543}{{\tt
  1104.4543}}].

\bibitem{tHooft2015}
G.~{'t Hooft}, {\it {Local conformal symmetry: The missing symmetry component
  for space and time}},  {\em International Journal of Modern Physics D} {\bf
  24} (2015), no.~12 1--5, [\href{http://www.arxiv.org/abs/1410.6675}{{\tt
  1410.6675}}].

\bibitem{Jizba}
P.~Jizba, L.~Rachwa{\l}, and J.~Kňap, {\it {Infrared behavior of Weyl gravity:
  Functional renormalization group approach}},  {\em Physical Review D} {\bf
  101} (feb, 2020) 044050, [\href{http://www.arxiv.org/abs/1912.10271}{{\tt
  1912.10271}}].

\bibitem{Salvio2018b}
A.~Salvio and A.~Strumia, {\it {Agravity up to infinite energy}},  {\em The
  European Physical Journal C} {\bf 78} (2018), no.~2 124,
  [\href{http://www.arxiv.org/abs/1705.03896}{{\tt 1705.03896}}].

\bibitem{Kugo1978-2}
T.~Kugo and I.~Ojima, {\it {Subsidiary conditions and physical S-matrix
  unitarity in indefinite-metric quantum gravitation theory}},  {\em Nuclear
  Physics B} {\bf 144} (nov, 1978) 234--252.

\bibitem{Nakanishi1990}
{Nakanishi, N. and Ojima, I.}, {\em {Covariant operator formalism of gauge
  theories and Quantum Gravity}}, vol.~27.
\newblock World Scientific, 1990.

\bibitem{Kugo}
T.~Kugo, {\em {Eichtheorie}}.
\newblock Springer Berlin Heidelberg, 1997.

\bibitem{Kawasaki1981}
S.~Kawasaki, T.~Kimura, and K.~Kitago, {\it {Canonical Quantum Theory of
  Gravitational Field with Higher Derivatives}},  {\em Progress of Theoretical
  Physics} {\bf 66} (dec, 1981) 2085--2102.

\bibitem{Kawasaki1982}
S.~Kawasaki and T.~Kimura, {\it {Canonical Quantum Theory of Gravitational
  Field with Higher Derivatives. II: Asymptotic Fields and Unitarity Problem}},
   {\em Progress of Theoretical Physics} {\bf 68} (nov, 1982) 1749--1764.

\bibitem{Kawasaki1983}
S.~Kawasaki and T.~Kimura, {\it {Canonical Quantum Theory of Gravitational
  Field with Higher Derivatives. III: A Formulation with an Additional BRS
  Invariance}},  {\em Progress of Theoretical Physics} {\bf 69} (mar, 1983)
  1015--1030.

\bibitem{Kugo1979b}
T.~Kugo and I.~Ojima, {\it {Local Covariant Operator Formalism of Non-Abelian
  Gauge Theories and Quark Confinement Problem}},  {\em Progress of Theoretical
  Physics Supplement} {\bf 66} (1979), no.~66 1--130.

\bibitem{Ostrogradsky1850}
M.~Ostrogradsky, {\it {M{\'{e}}moires sur les {\'{e}}quations
  diff{\'{e}}rentielles, relatives au probl{\`{e}}me des
  isop{\'{e}}rim{\`{e}}tres}},  {\em Mem. Acad. St. Petersbourg} {\bf 6}
  (1850), no.~4 385--517.

\bibitem{Woodard2015}
R.~P. Woodard, {\it {The Theorem of Ostrogradsky}},
  \href{http://www.arxiv.org/abs/1506.02210}{{\tt 1506.02210}}.

\bibitem{Boulware1983}
D.~G. Boulware, G.~T. Horowitz, and A.~Strominger, {\it {Zero-Energy Theorem
  for Scale-Invariant Gravity}},  {\em Physical Review Letters} {\bf 50} (may,
  1983) 1726--1729.

\bibitem{Salvio2016}
A.~Salvio and A.~Strumia, {\it {Quantum mechanics of 4-derivative theories}},
  {\em European Physical Journal C} {\bf 76} (2016), no.~4
  [\href{http://www.arxiv.org/abs/1512.01237}{{\tt 1512.01237}}].

\bibitem{Anselmi2018}
D.~Anselmi, {\it {Fakeons and Lee-Wick models}},  {\em Journal of High Energy
  Physics} {\bf 2018} (jan, 2018) 1--57,
  [\href{http://www.arxiv.org/abs/1801.00915}{{\tt 1801.00915}}].

\bibitem{Bender2008}
C.~M. Bender and P.~D. Mannheim, {\it {No-ghost theorem for the fourth-order
  derivative pais-uhlenbeck oscillator model}},  {\em Physical Review Letters}
  {\bf 100} (2008), no.~11 1--4,
  [\href{http://www.arxiv.org/abs/0706.0207}{{\tt 0706.0207}}].

\bibitem{Mannheim2018}
P.~D. Mannheim, {\it {Unitarity of loop diagrams for the ghost-like
  propagator}},  {\em arXiv} (2018), no.~5 1--18,
  [\href{http://www.arxiv.org/abs/1801.03220}{{\tt 1801.03220}}].

\bibitem{Salvio2019a}
A.~Salvio, {\it {Quasi-conformal models and the early universe}},  {\em The
  European Physical Journal C} {\bf 79} (sep, 2019) 750.

\bibitem{Salvio2021}
A.~Salvio, {\it {Dimensional transmutation in gravity and cosmology}},  {\em
  International Journal of Modern Physics A} {\bf 36} (mar, 2021) 2130006,
  [\href{http://www.arxiv.org/abs/2012.11608}{{\tt 2012.11608}}].

\bibitem{Donoghue2021}
J.~F. Donoghue and G.~Menezes, {\it {Ostrogradsky instability can be overcome
  by quantum physics}},  {\em Physical Review D} {\bf 104} (2021), no.~4 1--11,
  [\href{http://www.arxiv.org/abs/2105.00898}{{\tt 2105.00898}}].

\bibitem{Alvarez-Gaume2016}
L.~Alvarez-Gaume, A.~Kehagias, C.~Kounnas, D.~L{\"{u}}st, and A.~Riotto, {\it
  {Aspects of quadratic gravity}},  {\em Fortschritte der Physik} {\bf 64}
  (2016), no.~2-3 176--189, [\href{http://www.arxiv.org/abs/1505.07657}{{\tt
  1505.07657}}].

\bibitem{Kubo2022}
J.~Kubo and J.~Kuntz, {\it {Analysis of unitarity in conformal quantum
  gravity}},  {\em Classical and Quantum Gravity} (jul, 2022)
  [\href{http://www.arxiv.org/abs/2202.08298}{{\tt 2202.08298}}].

\bibitem{Riegert1984}
R.~J. Riegert, {\it {The particle content of linearized conformal gravity}},
  {\em Physics Letters A} {\bf 105} (1984), no.~3 110--112.

\bibitem{DeRham2014}
C.~de~Rham, {\it {Massive gravity}},  {\em Living Reviews in Relativity} {\bf
  17} (2014) 1--186, [\href{http://www.arxiv.org/abs/1401.4173}{{\tt
  1401.4173}}].

\bibitem{Hindawi1996}
A.~Hindawi, B.~A. Ovrut, and D.~Waldram, {\it {Consistent spin-two coupling and
  quadratic gravitation}},  {\em Physical Review D - Particles, Fields,
  Gravitation and Cosmology} {\bf 53} (1996), no.~10 5583--5596,
  [\href{http://www.arxiv.org/abs/9509142}{{\tt 9509142}}].

\bibitem{Salvio2019}
A.~Salvio, {\it {Metastability in quadratic gravity}},  {\em Physical Review D}
  {\bf 99} (2019), no.~10 [\href{http://www.arxiv.org/abs/1902.09557}{{\tt
  1902.09557}}].

\bibitem{Pottel2020a}
S.~Pottel and K.~Sibold, {\it {On the Perturbative Quantization of
  Einstein-Hilbert Gravity Embedded in a Higher Derivative Model}},  {\em
  Physical Review D} {\bf 104} (dec, 2020) 086012,
  [\href{http://www.arxiv.org/abs/2012.11450}{{\tt 2012.11450}}].

\bibitem{Becchi1975}
C.~Becchi, A.~Rouet, and R.~Stora, {\it {Renormalization of the abelian
  Higgs-Kibble model}},  {\em Communications in Mathematical Physics} {\bf 42}
  (1975), no.~2 127--162.

\bibitem{Becchi1976}
C.~Becchi, A.~Rouet, and R.~Stora, {\it {Renormalization of Gauge Theories}},
  {\em Annals of Physics} {\bf 98} (1976) 287--321.

\bibitem{Oda2022}
I.~Oda, {\it {Quantum theory of Weyl-invariant scalar-tensor gravity}},  {\em
  Physical Review D} {\bf 105} (2022), no.~12 1--37,
  [\href{http://www.arxiv.org/abs/2204.11200}{{\tt 2204.11200}}].

\bibitem{Rachwa2022}
L.~Rachwa{\l}, {\it {Introduction to Quantization of Conformal Gravity}},  {\em
  Universe} {\bf 8} (2022), no.~4 1--51,
  [\href{http://www.arxiv.org/abs/2204.13856}{{\tt 2204.13856}}].

\bibitem{Lehmann1955}
H.~Lehmann, K.~Symanzik, and W.~Zimmermann, {\it {Zur Formulierung
  quantisierter Feldtheorien}},  {\em Il Nuovo Cimento} {\bf 1} (1955), no.~1
  205--225.

\bibitem{Kugo1978-1}
T.~Kugo and I.~Ojima, {\it {Manifestly Covariant Canonical Formulation of the
  Yang-Mills Field Theories I: General Formalism}},  {\em Progress of
  Theoretical Physics} {\bf 60} (dec, 1978) 1869--1889.

\bibitem{Kugo1979a}
T.~Kugo and I.~Ojima, {\it {Manifestly Covariant Canonical Formulation of
  Yang-Mills Field Theories II: SU(2) Higgs-Kibble Model with Spontaneous
  Symmetry Breaking}},  {\em Progress of Theoretical Physics} {\bf 61} (jan,
  1979) 294--314.

\bibitem{Kugo1979}
T.~Kugo and I.~Ojima, {\it {Manifestly Covariant Canonical Formulation of
  Yang-Mills Field Theories III: Pure Yang-Mills Theories without Spontaneous
  Symmetry Breaking}},  {\em Progress of Theoretical Physics} {\bf 61} (feb,
  1979) 644--655.

\bibitem{Donoghue2019}
J.~F. Donoghue and G.~Menezes, {\it {Unitarity, stability, and loops of
  unstable ghosts}},  {\em Physical Review D} {\bf 100} (2019), no.~10
  [\href{http://www.arxiv.org/abs/1908.02416}{{\tt 1908.02416}}].

\bibitem{Veltman1963}
M.~Veltman, {\it {Unitarity and causality in a renormalizable field theory with
  unstable particles}},  {\em Physica} {\bf 29} (mar, 1963) 186--207.

\bibitem{Cutkosky1960}
R.~E. Cutkosky, {\it {Singularities and Discontinuities of Feynman
  Amplitudes}},  {\em Journal of Mathematical Physics} {\bf 1} (sep, 1960)
  429--433.

\bibitem{Denner2015}
A.~Denner and J.-N. Lang, {\it {The complex-mass scheme and unitarity in
  perturbative quantum field theory}},  {\em The European Physical Journal C}
  {\bf 75} (aug, 2015) 377, [\href{http://www.arxiv.org/abs/1406.6280}{{\tt
  1406.6280}}].

\bibitem{Lee1969a}
T.~D. Lee and G.~C. Wick, {\it {Negative metric and the unitarity of the
  S-matrix}},  {\em Nuclear Physics, Section B} {\bf 9} (1969), no.~2 209--243.

\bibitem{Martin-Garcia2008}
J.~M. Mart{\'{i}}n-Garc{\'{i}}a, {\it {xPerm: fast index canonicalization for
  tensor computer algebra}},  {\em Computer Physics Communications} {\bf 179}
  (2008), no.~8 597--603, [\href{http://www.arxiv.org/abs/0803.0862}{{\tt
  0803.0862}}].

\bibitem{Brizuela2009}
D.~Brizuela, J.~M. Mart{\'{i}}n-Garc{\'{i}}a, and G.~A. {Mena Marug{\'{a}}n},
  {\it {xPert: Computer algebra for metric perturbation theory}},  {\em General
  Relativity and Gravitation} {\bf 41} (2009), no.~10 2415--2431,
  [\href{http://www.arxiv.org/abs/0807.0824}{{\tt 0807.0824}}].

\bibitem{Nutma2014a}
T.~Nutma, {\it {XTras: A field-theory inspired xAct package for mathematica}},
  {\em Computer Physics Communications} {\bf 185} (2014), no.~6 1719--1738,
  [\href{http://www.arxiv.org/abs/1308.3493}{{\tt 1308.3493}}].

\bibitem{Frob2020}
M.~B. Fr{\"{o}}b, {\it {FieldsX -- An extension package for the xAct tensor
  computer algebra suite to include fermions, gauge fields and BRST
  cohomology}},  \href{http://www.arxiv.org/abs/2008.12422}{{\tt 2008.12422}}.

\bibitem{Pais1950}
A.~Pais and G.~E. Uhlenbeck, {\it {On Field Theories with Non-Localized
  Action}},  {\em Physical Review} {\bf 79} (jul, 1950) 145--165.

\bibitem{Ilhan2013}
I.~B. Ilhan and A.~Kovner, {\it {Some comments on ghosts and unitarity: The
  Pais-Uhlenbeck oscillator revisited}},  {\em Physical Review D - Particles,
  Fields, Gravitation and Cosmology} {\bf 88} (jan, 2013) 1--14,
  [\href{http://www.arxiv.org/abs/1301.4879}{{\tt 1301.4879}}].

\bibitem{Deffayet2021}
C.~Deffayet, S.~Mukohyama, and A.~Vikman, {\it {Ghosts without runaway}},
  \href{http://www.arxiv.org/abs/2108.06294}{{\tt 2108.06294}}.

\bibitem{Zimmermann1968}
W.~Zimmermann, {\it {The power counting theorem for Minkowski metric}},  {\em
  Communications in Mathematical Physics} {\bf 11} (mar, 1968) 1--8.

\bibitem{Hepp1966}
K.~Hepp, {\it {Proof of the Bogoliubov-Parasiuk theorem on renormalization}},
  {\em Communications in Mathematical Physics} {\bf 2} (dec, 1966) 301--326,
  [\href{http://www.arxiv.org/abs/2206.04072}{{\tt 2206.04072}}].

\bibitem{Appelquist1969}
T.~Appelquist, {\it {Parametric integral representations of renormalized
  feynman amplitudes}},  {\em Annals of Physics} {\bf 54} (1969), no.~1 27--61.

\end{thebibliography}\endgroup

\end{document}